\documentclass[conference]{IEEEtran}
\IEEEoverridecommandlockouts
\usepackage{cite}
\usepackage{amsmath,amssymb,amsfonts}
\usepackage{algorithmic}
\usepackage{graphicx}
\usepackage{textcomp}
\usepackage{xcolor}
\usepackage{subfigure} 
\usepackage{listings}
\usepackage{multirow}
\usepackage{tabularx}
\usepackage{adjustbox}
\usepackage{array}
\usepackage[normalem]{ulem}
\usepackage[linesnumbered,ruled]{algorithm2e}
\makeatletter
\let\NAT@parse\undefined
\makeatother
\useunder{\uline}{\ul}{}
\usepackage{tikz}
\def\BibTeX{{\rm B\kern-.05em{\sc i\kern-.025em b}\kern-.08em
    T\kern-.1667em\lower.7ex\hbox{E}\kern-.125emX}}

\usepackage[backref]{hyperref} 
\hypersetup{
hidelinks
}

\makeatletter
\newcommand{\linebreakand}{%
  \end{@IEEEauthorhalign}
  \hfill\mbox{}\par
  \mbox{}\hfill\begin{@IEEEauthorhalign}
}
\makeatother

\begin{document}

\title{\textbf{Firzen}: \textbf{Fir}ing Strict Cold-Start Items with Fro\textbf{zen} Heterogeneous and Homogeneous Graphs for Recommendation
}

\author{\IEEEauthorblockN{Hulingxiao He}
\IEEEauthorblockA{\textit{Wangxuan Institute of Computer Technology} \\ 
\textit{\& National Key Laboratory for Multimedia} \\ \textit{Information Processing, Peking University} \\
Beijing, China \\
hehulingxiao@stu.pku.edu.cn}
\and
\IEEEauthorblockN{Xiangteng He$^{\ast}$ \thanks{*Corresponding author.}}
\IEEEauthorblockA{\textit{Wangxuan Institute of Computer Technology} \\
\textit{\& National Key Laboratory for Multimedia} \\ \textit{Information Processing, Peking University} \\
Beijing, China \\
hexiangteng@pku.edu.cn}
\linebreakand
\IEEEauthorblockN{Yuxin Peng}
\IEEEauthorblockA{\textit{Wangxuan Institute of Computer Technology} \\
\textit{\& National Key Laboratory for Multimedia} \\ \textit{Information Processing, Peking University} \\
Beijing, China \\
pengyuxin@pku.edu.cn}
\and
\IEEEauthorblockN{Zifei Shan}
\IEEEauthorblockA{\textit{WeiXin Group} \\
\textit{Tencent}\\
Shanghai, China \\
zifeishan@tencent.com}
\and
\IEEEauthorblockN{Xin Su}
\IEEEauthorblockA{\textit{WeiXin Group} \\
\textit{Tencent}\\
Shenzhen, China \\
levisu@tencent.com}
}

\maketitle
\begin{abstract}
Recommendation models utilizing unique identities (IDs) to represent distinct users and items have dominated the recommender systems literature for over a decade. Since multi-modal content of items (e.g., texts and images) and knowledge graphs (KGs) may reflect the interaction-related users' preferences and items' characteristics, they have been utilized as useful side information to further improve the recommendation quality. However, the success of such methods often limits to either warm-start or strict cold-start item recommendation in which some items neither appear in the training data nor have any interactions in the test stage: (1) Some fail to learn the embedding of a strict cold-start item since side information is only utilized to enhance the warm-start ID representations; (2) The others deteriorate the performance of warm-start recommendation since unrelated multi-modal content or entities in KGs may blur the final representations. In this paper, we propose a unified framework incorporating multi-modal content of items and KGs to effectively solve both strict cold-start and warm-start recommendation termed \textit{Firzen}, which extracts the user-item collaborative information over frozen \textit{heterogeneous graph} (collaborative knowledge graph), and exploits the item-item semantic structures and user-user behavioral association over frozen \textit{homogeneous graphs} (item-item relation graph and user-user co-occurrence graph). Furthermore, we build four unified strict cold-start evaluation benchmarks based on publicly available Amazon datasets and a real-world industrial dataset from Weixin Channels via rearranging the interaction data and constructing KGs. Extensive empirical results demonstrate that our model yields significant improvements for strict cold-start recommendation and outperforms or matches the state-of-the-art performance in the warm-start scenario. The code is available at \href{https://github.com/PKU-ICST-MIPL/Firzen_ICDE2024}{\textcolor{blue}{https://github.com/PKU-ICST-MIPL/Firzen\_ICDE2024}}. 
\end{abstract}

\begin{IEEEkeywords}
Strict cold-start item recommendation, warm-start item recommendation, multi-modal recommendation, knowledge-aware recommendation
\end{IEEEkeywords}

\section{Introduction}

Recommender systems, which aim at suggesting items to users given historical user-item interactions, have been playing a crucial role for mitigating information overload in many online services, ranging from video-sharing cites\cite{liu2019uservideo}, online advertising\cite{gharibshah2021user} and E-commerce platforms\cite{wang2020time}. The modern recommendation models often use unique identities (IDs) to represent users and items, which are subsequently converted to embedding vectors as learnable parameters. These ID-based models have dominated the recommender system field for over a decade, especially in the warm-start scenarios when users and items have sufficient interaction data \cite{koren2009matrix, rendle2009bpr, he2020lightgcn, yuan2022tenrec}.

In recent years, to explore the rich multi-modal content of items and the fruitful facts contained in the knowledge graphs (KGs), some works have studied effective means to integrate such side information of items into the traditional user-item recommendation paradigm. On one hand, multi-modal content of items, such as visual and textual features of items, may reflect the items' characteristics and thus users' preferences from different perspectives \cite{wei2019mmgcn,zhou2023bootstrap, wei2023multi}. On the other hand, KGs, serving as useful external sources, can encode additional item-wise semantic association to enhance the user and item representations \cite{wang2019kgat, xian2019reinforcement, huang2021knowledge, zhao2017meta, wang2018ripplenet, wang2019explainable, wang2019kgat, tai2020mvin,wang2021learning,xia2021knowledge}. Moreover, \cite{sun2020multi} represents multi-modal content as nodes and integrate them into collaborative KGs for recommendation. 

\begin{figure}[t]
  \centering
  \includegraphics[width=0.9\linewidth]{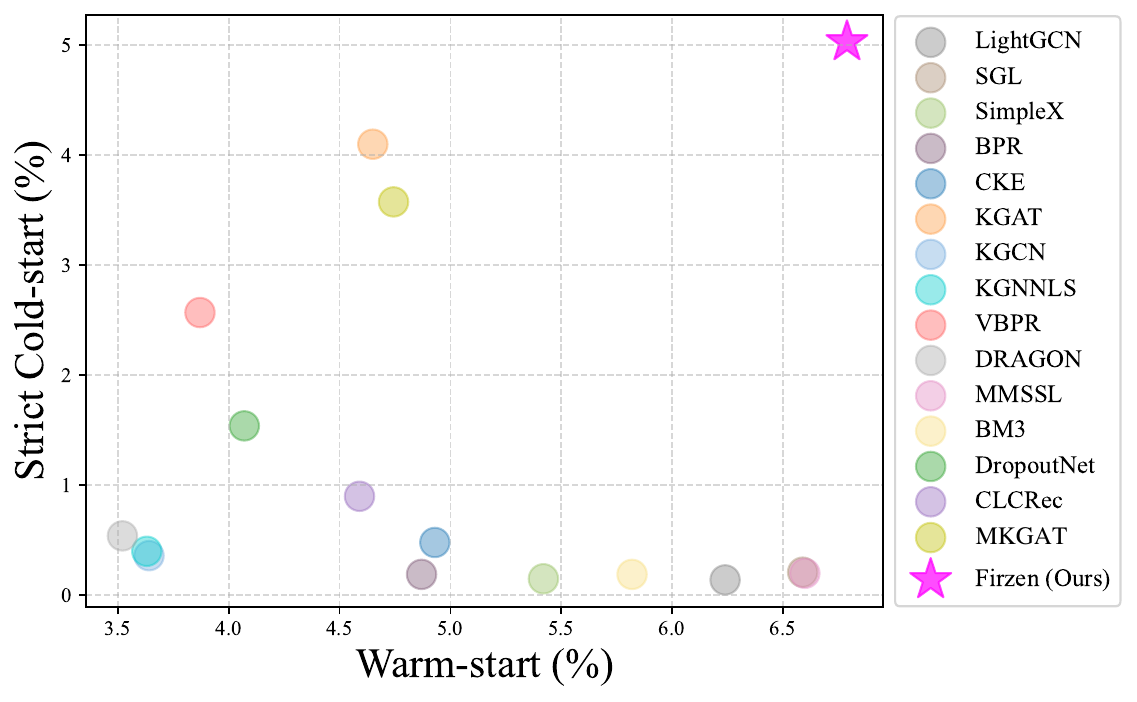}
   \caption{Performance comparison (MRR@20) of Firzen and existing methods on Amazon Beauty dataset for both strict cold-start and warm-start scenarios.}
   \label{fig:motivation1}
\end{figure}

\begin{figure}[t]
  \centering
  \includegraphics[width=0.99\linewidth]{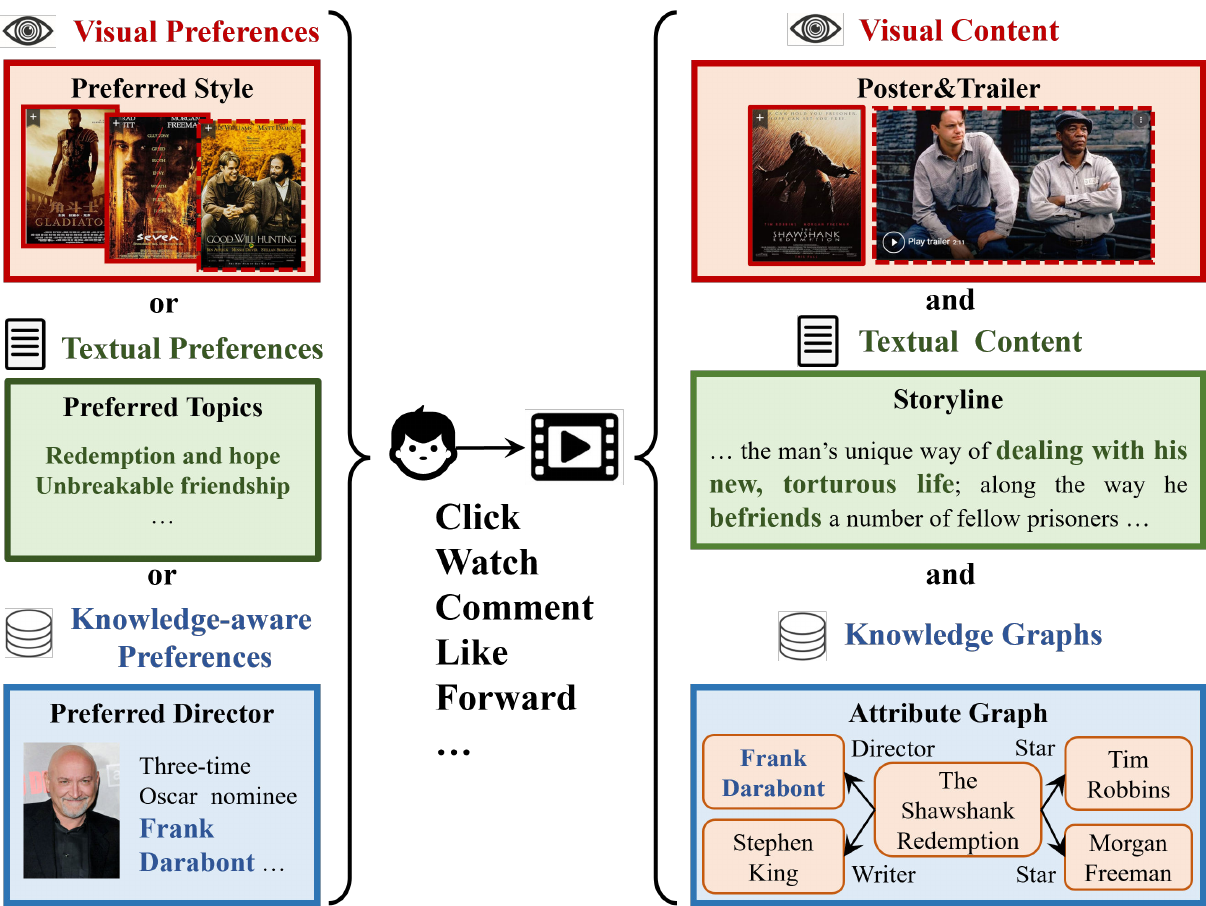}
   \caption{Illustrated example of users' preferences for items reflected on the visual content, textual content and knowledge graphs.}
   \label{fig:motivation2}
\end{figure}

As shown in Fig. \ref{fig:motivation1}, though existing recommendation methods incorporating either multi-modal content of items or KGs can achieve relatively superior accuracy, they are confronted with the trade-off issue. It prevents recommendation models from achieving state-of-the-art performance in both warm-start and strict cold-start item recommendation in which some items neither appear in the training data nor have any
interactions in the test stage. 
\textit{\textbf{(1) Strong dependence on interaction-based supervision:}} Some methods incorporate side information to enhance the learning of ID embeddings of warm-start items. However, strict cold-start items fail to learn feasible representations at the training phase since supervision (i.e., interaction data) is missing. Worse still, unlike normal cold-start items that have user-item link at the test phase, strict ones even cannot obtain information from well-trained user representations. Consequently, such methods (e.g., MMSSL, SGL) underperform in the strict cold-start scenario. \textit{\textbf{(2) Non-robustness to irrelevant content features:}} The others represent items from their content features which alleviate the dependency on the historical
interactions, benefiting the strict cold-start items. However, users’ preferences on items may be reflected on different side information. Incorporating all multi-modal content and entities in KGs irrelevant to interactions may hurt the performance of warm-start item recommendation. Thus, such methods (e.g., KGAT, VBPR) underperform in the warm-start scenario. For instance, as shown in Fig. \ref{fig:motivation2}, a user interacted with the movie 
\textit{The Shawshank Redemption} may be attracted by the movie poster, story line or just the director \textit{Frank Darabont}. 
As a consequence, \textbf{the failure to collaborate the interaction and content features in a feasible manner} brings an obstacle to balancing warm-start and strict cold-start.

In light of the aforementioned limitations and challenges, we develop a unified framework incorporating multi-modal content of items and KGs for handling both strict cold-start and warm-start item recommendation, termed \textbf{Firzen}, which involves \textbf{Side information-Aware Heterogeneous Graph Learning} (SAHGL) and \textbf{Modality-Specific Homogeneous Graph Learning} (MSHGL). More specifically, SAHGL exploits behavior-aware, modality-aware and knowledge-aware user and item representations on the constructed heterogeneous graph, i.e., collaborative knowledge graph. Different from \cite{volkovs2017dropoutnet}, the heterogeneous graphs are \textit{frozen} without random dropout during the training phase. MSHGL first builds the homogeneous graphs, including modality-specific item-item relation graphs according to the modality information and a user-user co-occurrence graph based on the behavioral records. Different from \cite{zhang2021mining}, the homogeneous graphs are \textit{frozen} without updating during the training phase. Then the information is propagated and aggregated among items and users respectively, to exploit the item-item semantic structure and user-user preference association. To evaluate models using multiple side information on strict cold-start item recommendation, we build four benchmarks via rearranging the raw interaction data and constructing KGs based on Amazon datasets and a real-world industrial dataset Weixin-Sports. Evaluation results show that Firzen can significantly improve the performance of strict cold-start recommendation while preserving competitive in warm-start scenario.  

To summarize, we make the following contributions:

\noindent (1) We propose a unified framework that can be trained effectively on warm-start items and seamlessly applied to strict cold-start items during inference phase.

\noindent (2) We develop a side information-aware user-item message passing mechanism based on frozen heterogeneous graph to encode the collaborative signals. We design a modality-specific item-item and user-user message passing mechanism based on frozen homogeneous graphs to exploit the internal relations within items and users, respectively.

\noindent (3) We build four strict cold-start benchmarks for evaluating recommendation methods using multiple side information. Extensive experiments on Amazon datasets and the real-world industrial dataset Weixin-Sports demonstrate that our proposed framework can significantly improve the performance on strict cold-start while preserving warm-start accuracy.

\section{PRELIMINARIES}
\begin{figure}[t]
    \subfigure[warm-start and normal item cold-start scenarios]{
       \centering
        \includegraphics[width=0.45\linewidth]{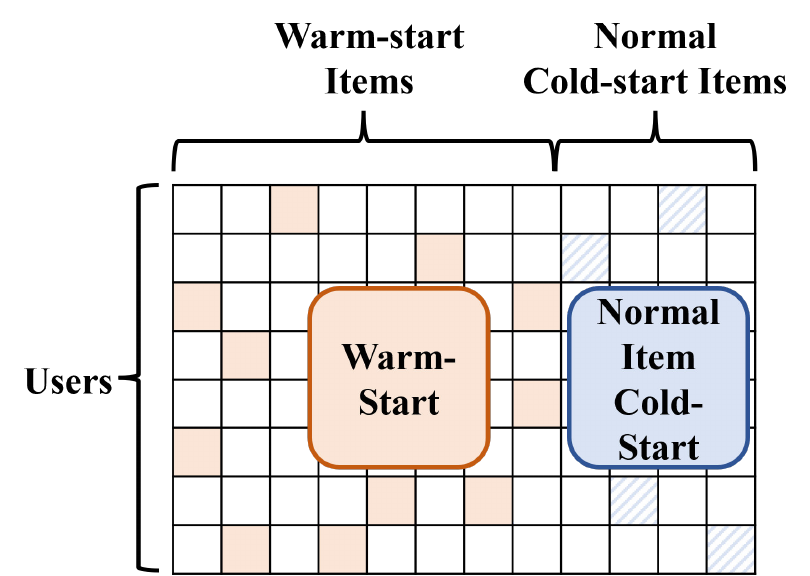}
    }
    \subfigure[warm-start and strict item cold-start scenarios]{
   \centering
    \includegraphics[width=0.45\linewidth]{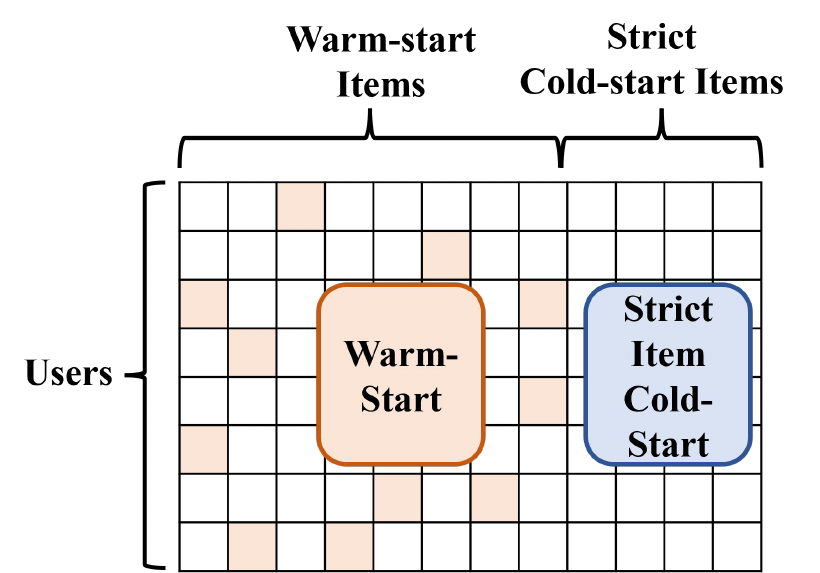}
    }    
    \caption{Warm-start, normal item cold-start and strict item cold-start scenarios. The filled and diagonal blocks indicate that there is an interaction record between the user and item at the training and inference phase, respectively.}
    \label{fig:normal&strict}
\end{figure}

Consider a recommendation dataset consisting of interaction records between a set $\mathcal{U}$ of users and a set $\mathcal{I}$ of items. We define the user-item interaction matrix $\mathcal{Y}$ to represent the interaction behaviors of users over different items (e.g., purchase, watch, review and click). In matrix $\mathcal{Y}$, the element $y_{u,i}=1$ given that the user $u$ has interacted with item $i$ and $y_{i,j}=0$, otherwise. The goal of classic \textit{warm-start recommendation} is to predict preferred items exist in the original interaction matrix $\mathcal{Y}$ for users seen during training, while the \textit{normal cold-start item recommendation} refers to predicting the preferred items unseen during training but have interactions at the test stage \cite{qian2020attribute}, as shown in Fig. \ref{fig:normal&strict} (a).

\textbf{Strict cold-start} is an extreme scenario of the normal cold-start, i.e., the strict cold-start items that neither appear in the training data nor have any interactions at the test stage \cite{qian2020attribute}, as shown in Fig. \ref{fig:normal&strict} (b). The significant differences lie on \textbf{whether interaction records exist at the test stage}, where normal cold-start items can explicitly link to users and thus can obtain information from well-trained user embeddings to obtain feasible representations. 

\textbf{User-Item Interaction Graph.} Based on the matrix $\mathcal{Y}$, we first construct the user-item interaction graph $\mathcal{G}_{inter}=\{\mathcal{V}_{inter}, \mathcal{E}_{inter}\}$, where the node set $\mathcal{V}_{inter} = \mathcal{U} \cup \mathcal{I}$ and edge $(u, i)$ is generated if $y_{u,i}=1$.

\textbf{Knowledge Graph.} Let $\mathcal{G}_{know}=\{(h,r,t)|h,t\in\mathcal{V}_{know},r\in\mathcal{E}_{know}\}$ represent the knowledge graph which organizes external item attributes with different types of entities $\mathcal{V}_{know}$ and relationships $\mathcal{E}_{know}$. Specifically, each triplet $(h, r, t)$ characterizes the semantic association between the head entity $h$ and the tail entity $t$. Such information incorporates fruitful external facts associated with items and connections among items to improve the modeling of users' underlying preferences for recommendation.

\textbf{Task Formulation.} We formally describe our task as follows. Input: user-item interaction data $\mathcal{G}_{inter}=\{\mathcal{V}, \mathcal{E}\}$, item knowledge graph data $\mathcal{G}_{know}=\{(h,r,t)\}$, and multi-modal data of items $\mathcal{F}_{I}$. Output: the learned unified function $Fun =(u,v|\mathcal{G}_{inter},\mathcal{G}_{know},\mathcal{F}_{I};\theta$) that forecasts the warm-start items from $\mathcal{I}_{warm}$ and strict cold-start items from $\mathcal{I}_{cold}$ that user $u \in \mathcal{U}$ would like to interact with, respectively. $\mathcal{\theta}$ denotes the model parameters. 

\section{METHODOLOGY}
\label{sec:methodology}
\subsection{Overview}

\begin{figure*}[htbp]
  \centering
  \includegraphics[width=\linewidth]{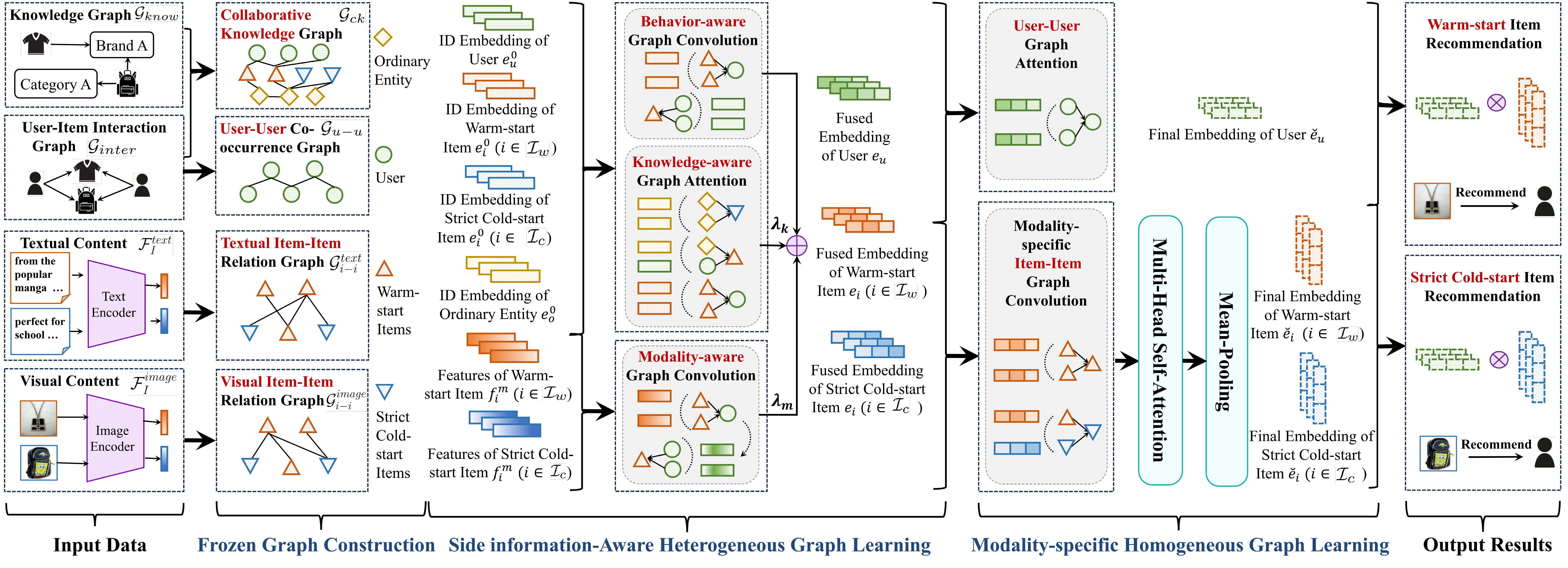}
   \caption{The model flow of the proposed Firzen architecture. The KGs, user-item interaction graph and multi-modal content are utilized to construct frozen graphs. Then the graphs together with the initial user/item/entity ID embeddings and multi-modal features are fed into SAHGL and MSHGL in succession to obtain final representations. $\lambda_{k}, \lambda_{m}$ represent the importance of knowledge-aware and modality-aware representations, respectively.}
   \label{fig:method}
\end{figure*}

Our proposed Firzen is a unified learning framework collaborating multi-modal content and KGs to improve the performance of strict cold-start item recommendation while maintaining competitive results on warm-start item recommendation. Fig. \ref{fig:method} depicts the overall model flow of Firzen which consists of three main components. The first component is Frozen Graph Construction, a pre-processing component to build the frozen graphs utilized in the following components. The second component is Side information-Aware Heterogeneous Graph Learning (SAHGL), a component for extracting user-item collaborative signals from behaviors, multi-modal content and KGs, respectively.  The third component is Modality-Specific Homogeneous Graph Learning (MSHGL), a component for propagating information from warm-start to both strict cold-start and warm-start items based on the item-item semantic structures of different modalities, and among users based on the user-user co-occurrence relation.

\subsection{Frozen Graph Construction}
\label{sec:graph_construct}
\textit{1) Collaborative Knowledge Graph.} To facilitate modeling behavior-aware and knowledge-aware user/item representations, we follow the definition in \cite{wang2019kgat} to build collaborative knowledge graph that encodes user behaviors and item knowledge as a unified relational graph. First, each edge in the user-item interaction graph $\mathcal{G}_{inter}$ is represented as a triplet $(u, Interact, i)$. Based on the item-entity alignment, $\mathcal{G}_{inter}$ is integrated with $\mathcal{G}_{know}$ as a unified collaborative knowledge graph $\mathcal{G}_{ck}=\{(h,r,t)|h,t\in \mathcal{V^{'}}, r\in \mathcal{E}^{'}\}$, where $\mathcal{V}^{'} = \mathcal{V}_{know} \cup \mathcal{V}_{inter}$ and $\mathcal{E}^{'} = \mathcal{E}_{know} \cup {Interact}$.

\textit{2) Modality-Specific Item-Item Relation Graph.} Motivated by the success of item-item graph in multi-modal recommendation \cite{zhang2021mining, zhou2022tale, zhou2023enhancing}, we propose to introduce item-item graphs to  bridge the gap between strict cold-start and warm-start items while preserving the performance of warm-start recommendation. The advantages are two-folds. On one hand, it can exploit the latent item content semantic from the item-item structure based on the similarities of items' multi-modal features. On the other hand, it can propagate the collaborative signals from warm-start items to strict cold-start items, boosting the recommendation performance of strict cold-start scenario.  

As different side information can reflect different characteristics of items, we construct modality-specific item-item relation graph represented by the adjacency matrix $G_{i-i}^{m}$ for each modality $m$. Specifically, we calculate the cosine similarities $G_{i-i}^{m}(a,b)$ for $a$-th item and $b$-th item on their pre-calculated raw features of side information $f_{a}^{m}$ and $f_{b}^{m}$:
\begin{equation}
    G_{i-i}^{m}(a,b) = \frac{f_{a}^{m}(f_{b}^{m})^{T}}{||f_{a}^{m}||_{2}||f_{b}^{m}||_{2}}.
\end{equation}
Then we utilize kNN sparsification \cite{chen2009fast} to convert similarity matrix $G_{i-i}^{m}$ into an unweighted matrix. For each item $a$, we only keep the top-K similar items connected to $a$:
\begin{equation}
    \tilde{G} _{i-i}^{m}(a, b) = \begin{cases} & 1,\text{ if } G _{i-i}^{m}(a, b) \in topK(G_{i-i}^{m}(a)) \\  &0, 
    otherwise.\end{cases} 
\end{equation}
where each element is either 0 or 1, with 1 denoting the semantic connection between the two items. We then normalize $\tilde{G} _{i-i}^{m}$ to build the final adjacency matrix $\Check{G}_{i-i}^{m}$:
\begin{equation}
    \Check{G}_{i-i}^{m}=(D^{m})^{-\frac{1}{2}}\tilde{G} _{i-i}^{m}(D^{m})^{-\frac{1}{2}},
\end{equation}
where $D^{m}$ denotes the diagonal degree matrix of $\tilde{G} _{i-i}^{m}$. We consider textual and visual modalities $M=\{text, image\}$ in our paper and then we obtain textual-based item-item relation graph $\Check{\mathcal{G}}_{i-i}^{text}$ and visual-based item-item relation graph $\Check{\mathcal{G}}_{i-i}^{image}$. In the training phase, the item-item graph is built on all warm-start items, while in the inference phase, the item-item graph is expanded to built on all items, including warm-start items and strict cold-start items.

\textit{3) User-User Co-occurrence Graph.} As users who have interacted with similar items often have similar preferences, we also build user-user co-occurrence graph represented by the adjacency matrix $G_{u-u}$ to learn the internal relations between users. Specifically, for each user $u \in \mathcal{U}$, we retain top-K users with the highest number of commonly interacted items \cite{zhou2023enhancing}:
\begin{equation}
\resizebox{0.9\linewidth}{!}{$
    \tilde{G}_{u-u}(a, b) = \begin{cases} & G_{u-u}(a, b),\text{ if } G_{u-u}(a, b) \in topK(G_{u-u}(a)) \\  &0, 
    otherwise.\end{cases} 
$}
\end{equation}
where $G_{u-u}(a,b)$ denotes the number of commonly interacted items for user $a$ and $b$. 

\subsection{Side information-Aware Heterogeneous Graph Learning}
\label{sec:SIH}
Inspired by the relative superiority of recommendation models incorporating side information in strict cold-start scenario and general collaborative filtering models in warm-start scenario, we first design three modules to encode behavior-aware, knowledge-aware and modality-aware user preferences and item content, respectively

\textit{1) Behavior-aware Graph Convolution.} 
Since graph-based collaborative filtering models (e.g., LightGCN\cite{he2020lightgcn}) can perform well in warm-start item recommendation, we first build our encoder upon the graph neural network for recursive message passing over the user-item interaction graph $\mathcal{G}_{inter}$ to explore the high-order collaborative effects without incorporating items' multiple side information. For layer $l$, the user and item embeddings are formulated as follows:
\begin{equation}
    e_{u}^{l+1}=\sum_{i\in \mathcal{N}_{u}}\frac{e_{i}^{l}}{\sqrt[]{|\mathcal{N}_{u} |} },
\end{equation}
\begin{equation}
    e_{i}^{l+1}=\sum_{u\in \mathcal{N}_{i}}\frac{e_{u}^{l}}{\sqrt[]{|\mathcal{N}_{i} |} }, 
\end{equation}
where $\mathcal{N}_{u}, \mathcal{N}_{i}$ denote the neighborhood sets of user $u$ and item $i$ in $\mathcal{G}_{inter}$, respectively. We initialize $e_{u}^{0}$ and $e_{i}^{0}$ by Xaiver uniform initialization. In the multi-layer GNNs, we utilize mean-pooling to aggregate layer-wise embeddings and obtain the final behavior-aware user embeddings: $\tilde{e}_{u}=\sum_{l=0}^{L} \frac{e_{u}^{l}}{L}$, and item embeddings: $\tilde{e}_{i}=\sum_{l=0}^{L} \frac{e_{i}^{l}}{L}$, where $L$ denotes the number of GNN layers. Note that there is no edge connected to strict cold-start items in $\mathcal{G}_{inter}$, and thus after the behavior-aware graph convolution, the embeddings of strict cold-start items are zero vectors, same as skipping the collaborative filtering module based on interaction record. 

\textit{2) Modality-aware Graph Convolution.} In order to excavate modality-specific user preferences and item characteristics, we first project raw multi-modal features into the interaction-related multi-modal features and then aggregate over user-item interactions, computed as:
\begin{equation}
    x_{u}^{m}=\sum_{i\in \mathcal{N}_{u}}\frac{Linear(f_{i}^{m})}{\sqrt[]{|\mathcal{N}_{u} |} },
\end{equation}
                    
\begin{equation}
    x_{i}^{m}=\sum_{u\in \mathcal{N}_{i}}\frac{x_{u}^{m}}{\sqrt[]{|\mathcal{N}_{i} |} },
\end{equation}
where $Linear(\cdot)$ is a fully-connected layer with dropout \cite{hinton2012improving} to project items' raw multi-modal features into the interaction-aware vector space. 

\textit{3) Knowledge-aware Graph Attention.} We build our message aggregation mechanism among users, items and the connected entities in $\mathcal{G}_{know}$, for generating knowledge-aware user embeddings and item embeddings based on the heterogeneous attentive aggregator \cite{wang2019kgat}. Specifically, we employ knowledge-aware attention to aggregate the neighborhood information on $\mathcal{G}_{ck}$, shown as follows:
\begin{equation}
    x_{\mathcal{N}_{h}} = \sum_{(h, r, t)\in\mathcal{N}_{h}}\alpha(h,r,t)x_{t},  
\end{equation}
\begin{equation}
    \alpha(h,r,t)= \frac{exp(\pi(h,r,t))}{\sum_{(h,r^{'},t^{'})\in\mathcal{N}_{h}}exp(\pi(h, r^{'}, t^{'}))},
\end{equation}
\begin{equation}
    \pi(h,r,t)= (W_{r}x_{t})^{T}tanh(W_{r}x_{h}+x_{r})), 
\end{equation}
where $\mathcal{N}_{h}=\{(h,r,t)|(h,r,t)\in\mathcal{G}_{ck}\}$ denotes the ego-network\cite{qiu2018deepinf} consisting of a head entity $h$ and the set of triplets connected to $h$. We initialize $x_h$ and $x_t$ by Xaiver uniform initialization. Note that $h$ can be arbitrary entity in $\mathcal{G}_{ck}$, including users, items and ordinary entities incorporated by the external KG $\mathcal{G}_{know}$. When $h$ represents user $u$ or item $i$, the relationship between the head entity embedding $x_{h}$ and initialized user/item id embedding $e^{0}_{u}$ and $e^{0}_{i}$ is summarized as:
\begin{equation}
 x_{h}=\left\{\begin{array}{ll}
e^{0}_{u}, & \text { when } h=u \in \mathcal{U} \\
e^{0}_{i}, & \text { when } h=i \in \mathcal{I}, \\
\end{array}\right.   
\end{equation}
\noindent After obtaining the ego-network representations $x_{\mathcal{N}_{h}}$, we then aggregate $x_{\mathcal{N}_{h}}$ with the entity representations via the bi-interaction aggregator:
\begin{equation}
\begin{split}
        x_{h}^{know} = LeakyReLU(W_{1}(x_{h}+x_{\mathcal{N}_{h}}))+ \\ LeakyReLU(W_{2}(x_{h}\odot x_{\mathcal{N}_{h}}))
\end{split}
\end{equation}
where $W_{1}, W_{2}$ denote the trainable weight matrices and $\odot$ is the element-wise product.

\textit{4) Importance-aware Information Fusion.} After obtaining behavior-aware, knowledge-aware and modality-aware user/item embeddings, we then fuse them to obtain the overall representations, which can reflect the interaction-related user preference and item characteristics based on external knowledge and different modalities:
\begin{equation}
    e_{u} = \tilde{e}_{u} + \lambda_{k} \cdot x_{u}^{know} +  \lambda_{m} \cdot (\beta_{t} \cdot x_{u}^{text} +  \beta_{i} \cdot x_{u}^{image}),
\end{equation}
\begin{equation}
    e_{i} = \tilde{e}_{i} + \lambda_{k} \cdot x_{i}^{know} +  \lambda_{m} \cdot (\beta_{t} \cdot x_{i}^{text} + \beta_{i} \cdot x_{i}^{image}),
\end{equation}
where $\lambda_{k}, \lambda_{m}$ control the ratio of behavior-aware, knowledge-aware and modality-aware representations. $\beta_{t}$ and $\beta_{i}$ denote the relative importance of textual and visual information that are optimized during the training phase.

The importance of different modalities lies on the contribution to capturing and predicting users' preferences and behaviors, and thus we propose to update the weights according to difficulty of distinguishing the generated virtual user-item interaction graph based on multi-modal content and the observed user-item interaction graph. Specifically, we collect the outputs of the discriminator $D(\cdot)$ that will be described in detail in Section \ref{sec:training_obj} when inputs are generated user-item interaction graph based on textual features $\Bar{\mathcal{G}}_{inter}^{text}$ and visual features $\Bar{\mathcal{G}}_{inter}^{image}$, respectively. Then, the output scores are normalized with softmax operation and utilized to update the importance score in a momentum manner:
\begin{equation}
    \beta_{t} = \eta \beta_{t} + (1 - \eta) \frac{exp(D(\Bar{\mathcal{G}}_{inter}^{text}))}{exp(D(\Bar{\mathcal{G}}_{inter}^{text}))+exp(D(\Bar{\mathcal{G}}_{inter}^{image}))},
\end{equation}
\begin{equation}
    \beta_{i} = \eta \beta_{i} + (1 - \eta) \frac{exp(D(\Bar{\mathcal{G}}_{inter}^{image}))}{exp(D(\Bar{\mathcal{G}}_{inter}^{text}))+exp(D(\Bar{\mathcal{G}}_{inter}^{image}))},
\end{equation}
where $\eta$ denotes the momentum to update $\beta_{t}$ and $\beta_{i}$.

\subsection{Modality-Specific Homogeneous Graph Learning}
\label{sec:MSH}

\textit{1) Message Passing on Homogeneous Graphs.} 
Based on the pre-constructed latent item-item relation graph of different modalities in Section \ref{sec:graph_construct} and the fused features of users/items in Section \ref{sec:SIH}, we utilize a light-weighted GCN \cite{he2020lightgcn} for information propagation and aggregation on $\Check{\mathcal{G}}_{i-i}^{text}$ and $\Check{\mathcal{G}}_{i-i}^{image}$ \cite{zhou2022tale}. Specifically, the graph convolution over the item-item relation graph is calculated as :
\begin{equation}
    \hat{h}_{l}^{m}(a) = \sum_{b\in \mathcal{N}_{a}}\Check{G}_{i-i}^{m}(a,b)\hat{h}_{l-1}^{m}(b),
\end{equation}
where $\hat{h}_{l}^{m}(a)$ denotes the representation of $a$-th item from the $l$-th layer, while $\hat{h}_{l-1}^{m}(b)$ denotes the representation of $b$-th item from the $l-1$-th layer. $\mathcal{N}(a)$ denotes the neighbor items of $a$-th item. $\Check{G}_{i-i}^{m}$ is the adjacency matrix of $\Check{\mathcal{G}}_{i-i}^{m}$. Note that $\hat{h}_{0}^{m}(a)=e_{a}$, which is the fused embedding obtained by SAHGL in Section \ref{sec:SIH}. After stacking $L_{i-i}$ convolutional layers on the item-item graphs, we then obtain the representations of items after modality-specific message passing $\hat{h}^{m}=\hat{h}_{L_{i-i}}^{m}$.

As neighbors with a higher number of commonly interacted items should have a larger effect on the user, the graph attention is designed for message passing on the user-user co-occurrence graph $\mathcal{G}_{u-u}$:
\begin{equation}
    \hat{z}_{l}(a) = \sum_{b\in \mathcal{N}_{a}}\frac{exp(\tilde{G}_{u-u}(a,b))}{\sum_{c\in\mathcal{N}_{a}}exp(\tilde{G}_{u-u}(a,c))}\hat{z}_{l-1}(b),
\end{equation}
where $\hat{z}_{l}(a)$ denotes the representation of $a$-th user from the $l$-th layer, while $\hat{z}_{l-1}(b)$ denotes the representation of $b$-th user from the $l-1$-th layer. $\mathcal{N}_{a}$ denotes the neighbor users of $a$-th user. $\tilde{G}_{u-u}$ is the adjacency matrix of $\tilde{\mathcal{G}}_{u-u}$. After stacking $L_{u-u}$ attention layers on the user-user graph, we then obtain final representations of users after message passing $\Breve{e}_{u}=\hat{z}_{L_{u-u}}$.

\textit{2) Dependency-aware Information Fusion.} To capture the correlations of items' representations extracted from modality-specific item-item relation graph, we employ a multi-head self-attention mechanism \cite{vaswani2017attention} along with mean-pooling to fuse the information aware of the modality dependency:
\begin{equation}
\resizebox{0.87\linewidth}{!}{$
    \hat{e}_{i}^{m}=\sum_{m'\in\mathcal{M}}\left |  \right |_{head=1}^{H}\sigma (\frac{e_{i}^{m}W_{head}^{Q}\cdot (e_{i}^{m'}W_{head}^{K})^{T} }{\sqrt{d/H} })\cdot e_{i}^{m'}, 
$}
\end{equation}
\begin{equation}
    \Breve{e}_{i} = \frac{1}{|\mathcal{M}|}\sum_{m\in\mathcal{M}}{\hat{e}_{i}^{m}},
\end{equation}
where $H$ denotes the number of attention heads, $\sigma(\cdot)$ denotes the softmax function. $W_{head}^{Q}, W_{head}^{K}\in\mathbb{R}^{d\times{d/H}}$ denote the $head$-th transformation matrix for query and key, respectively.

\subsection{Optimization}
\label{sec:training_obj}

\textit{1) Training Objectives.} To optimize the recommendation model, we opt for a multi-task training scheme as followed.

\textit{i) Adversarial Loss for Learning Interaction-related Multi-modal Content.} As the raw multi-modal content of items may contain irrelevant features than couldn't reflect the characteristics of items, we employ an adversarial self-supervised learning task to obtain the interaction-related modality content\cite{wei2023multi}. Specifically, the virtual interaction graph represented in the matrix form  $\Bar{G}_{inter}^{m}$ is constructed by the transformed multi-modal features $x_{u}^{m}$, $x_{i}^{m}$:
\begin{equation}
    \Bar{G}_{inter}^{m} = \frac{x_{u}^{m}\cdot (x_{i}^{m})^{T}}{||x_{u}^{m}||_{2}\cdot ||x_{i}^{m}||_{2}}.
\end{equation}
Instead of taking the observed user-item interaction graph, the objective graph for reconstruction is augmented by Gumbel-Softmax\cite{jang2017categorical} and the auxiliary signals with the final user and item embeddings $\Breve{e}_{u}$ and $\Breve{e}_{i}$:
\begin{equation}
    G_{inter}^{aug}(a, b) = \frac{exp((G_{inter}(a,b)+g)/\tau)}{\sum_{b'}{exp((G_{inter}(a,b')+g)/\tau)}} + \gamma \cdot \phi,
\end{equation}
\begin{equation}
    \phi  = \frac{\Breve{e}_{u}\Breve{e}_{i}}{||\Breve{e}_{u}||_{2}||\Breve{e}_{i}||_{2}},
\end{equation}
\begin{equation}
    g = -log(-log(uni)), uni\sim Uniform(0,1), 
\end{equation}
where $\tau$ is the temperature factor and $\gamma$ is a weight parameter to control the auxiliary signals. With the Gumbel-based transformation\cite{jang2017categorical}, we obtain the objective user-item graph $\mathcal{G}_{inter}^{aug}$. An auxiliary discriminator $D(x)=sigmoid(Drop(BN(LeakyReLU(Linear(x)))))$ is introduced to distinguish $\mathcal{G}_{inter}^{aug}$ and $\Bar{\mathcal{G}}_{inter}^{m}$, and the optimization loss is defined as follows:
\begin{equation}
    L_{adv} = \mathbb{E}_{\mathcal{G}_{inter}^{aug}}[D(\mathcal{G}_{inter}^{aug})]-\mathbb{E}_{\Bar{\mathcal{G}}_{inter}^{m}}[D(\Bar{\mathcal{G}}_{inter}^{m})]+\xi \cdot p,
\end{equation}
\begin{equation}
    p = \mathbb{E}_{\mathcal{G}_{inter}^{*}}[(||\nabla_{D(\mathcal{G}_{inter}^{*})}||-1)^{2}],
\end{equation}
where $\mathcal{G}_{inter}^{*}$ denotes the interpolation of $\mathcal{G}_{inter}^{aug}$ and $\Bar{\mathcal{G}}_{inter}^{m}$, and $p$ denotes the gradient penalty introduced from WassersteinGAN-GP \cite{gulrajani2017improved} with weight $\xi$.

\textit{ii) Contrastive Loss for Learning Diverse Modality-specific User Preferences.} The InfoNCE loss \cite{he2020momentum} is employed to maximize the mutual information between the modality-aware user embeddings $x_{u}^{m}$ and the final user embeddings $\Breve{e}_{u}$:
\begin{equation}
    L_{contr} = -\sum_{m\in\mathcal{M} }\sum_{u\in\mathcal{U}}log\frac{exp\,s(\breve{e}_{u}, x_{u}^{m} )}{den},
\end{equation}
\begin{equation}
    den = {\sum_{u'\in \mathcal{U}}(exp\,s(\breve{e}_{u'},x_{u}^{m})+exp\,s(e_{u'}^{m},x_{u}^{m}))  {} },
\end{equation}
where $s$ denotes the cosine similarity function.

\textit{iii) Knowledge Graph Representation Loss for Learning the Triplet Semantics.} Since we introduce the external KGs, we follow TransR \cite{lin2015learning} to encourage the discrimination of valid and broken triplets in KGs through a pairwise ranking loss:
\begin{equation}
\resizebox{0.88\linewidth}{!}{$
    L_{KG}=\sum_{(h,r,t_{p},t_{n})\in \kappa}-ln\,sigmoid(sc(h,r,t_{p})-sc(h,r,t_{n}))),
    $}
\end{equation}
\begin{equation}
    sc(h,r,t) = -||W_{r}e_{h}+e_{r}-W_{r}e_{t}||_{2}^{2},
\end{equation}
where $\kappa = \{(h,r,t_{p},t_{n})|(h,r,t_{p})\in \mathcal{G}_{know},(h,r,t_{n})\notin \mathcal{G}_{know}\}$, and $W_{r}$ is the transformation matrix of relation $r$. 

\textit{iv) Multi-task Training.} 
By integrating the \textit{side information-aware heterogeneous graph learning} and \textit{modality-specific homogeneous graph learning} components together, the training for recommendation is implemented by optimizing multiple objectives jointly:
\begin{equation}
    L_{Rec} = L_{BPR} + \lambda_{adv}  L_{adv} + \lambda_{contr}  L_{contr} + \lambda_{reg} ||\theta||^{2},
\end{equation}
\begin{equation}
    L_{BPR} = \sum_{(u,i_{p},i_n)}{-log(sigmoid(\hat{y}_{u,i_{p}}-\hat{y}_{u,i_{n}}))},
\end{equation}
where $(u,i_{p},i_{n})$ denotes a triplet comprised of user, positive item and negative item, $\lambda_{adv}$, $\lambda_{contr}$ and $\lambda_{reg}$ control the weights of different loss terms.

We optimize $L_{KG}$ and $L_{Rec}$ alternatively during a training step to improve the knowledge graph representations and recommendation performance.

\textit{2) Time Complexity Analysis.} The time cost mainly comes from the following parts. i) For calculating adversarial loss, it takes $O(|\mathcal{I}|\sum_{m\in\mathcal{M}}d^{m}d)$, $O(|\mathcal{M}|B|\mathcal{I}|d)$ and $O(B|\mathcal{I}|)$ for generating interaction-related multi-modal content, constructing user-item interaction graph based on transformed features and discriminating by $D(\cdot)$, where $B$ represents the training batch size. ii) For calculating the contrastive loss, it takes $O(|\mathcal{M}||\mathcal{U}|Bd)$. iii) The KG representation loss and BPR loss require $O(B(d^{know})^{2})$ and $O(Bd)$, respectively.

\subsection{Inference Procedure}
\label{sec:inference}
Our Firzen framework can be seamlessly applied to both warm-start and strict cold-start items during the inference phase. The only difference between training and inference procedure is an extra mask $M_{i-i}^{m}$ to isolate information propagation from strict cold-start items to warm-start items:
\begin{equation}
    M_{i-i}^{m}(a, b) = \begin{cases} & 0,\text{ if } a \in  \mathcal{I}_{warm}  \text{ and }  b \in \mathcal{I}_{cold}\\  &1, \text{ otherwise}.\end{cases} 
\end{equation}
The modality-specific item-item relation graph is rectified as:
\begin{equation}
    \hat{G}_{i-i}^{m} = \tilde{G} _{i-i}^{m} \odot M_{i-i}^{m},
\end{equation}
where the $\odot$ denotes the element-wise product of two matrices.

\subsection{Implementation Details}
\label{sec:implementation}
We implement Firzen using Pytorch 1.13.1 and train on a 64-bit Linux server equipped with 48 Intel Xeon@2.20GHz CPUs, 256GB memory, and 4 TITAN Xp GPUs. We choose 64 as the embedding dimension. The batch size is set to 2048. We choose the Adam optimizer \cite{kingma2015adam} to train for 300 epochs and perform early stopping. 

\section{EXPERIMENTS}
To evaluate the performance of our proposed Firzen framework, we conduct extensive experiments to answer the following research questions:
\begin{itemize}
    \item \textbf{RQ1} Whether Firzen outperforms the existing methods on both strict cold-start and warm-start recommendation?
    \item \textbf{RQ2} How does each component of Firzen contribute?
    \item \textbf{RQ3} How do the hyperparameters influence?
    \item \textbf{RQ4} How effective is Firzen in alleviating KG noise issues for recommendation?
    \item \textbf{RQ5} How effective is Firzen transferred to normal cold-start item recommendation?
    \item \textbf{RQ6} How are the training and inference time affected?
    \item \textbf{RQ7} How to explain the interpretability of Firzen?
    \item \textbf{RQ8} Where do the improvements of Firzen come from?
\end{itemize}

\subsection{Experimental Setup}

\textit{1) Datasets:} We use three publicly available recommendation datasets from Amazon\footnote{http://jmcauley.ucsd.edu/data/amazon/links.html} and a real-world industrial dataset Weixin-Sports collected from Weixin Channels. We rearrange them to build four unified strict cold-start evaluation benchmarks for models utilizing different side information. 

\begin{figure}[t]
   \centering
    \includegraphics[width=0.75\linewidth]{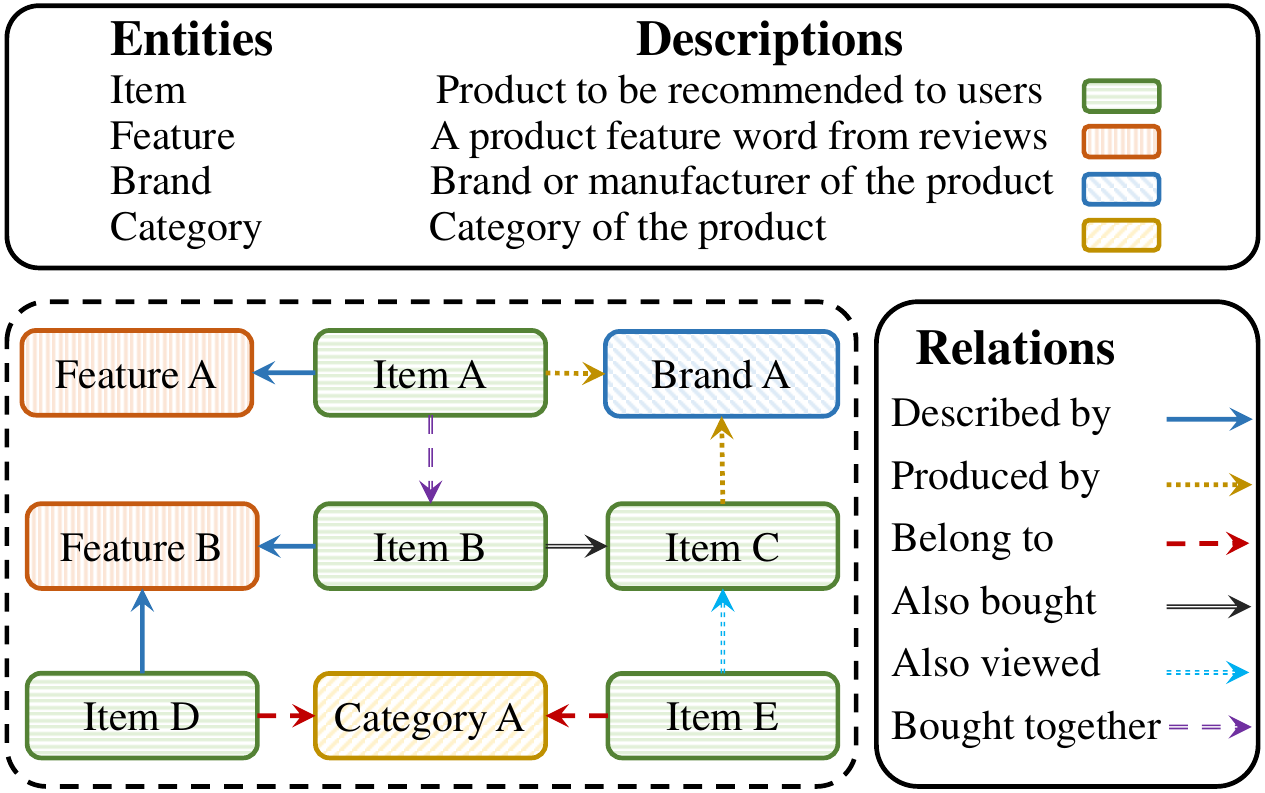}
     \caption{Illustration of the constructed knowledge graphs for Amazon datasets.}
    \label{fig:kg}
\end{figure}

\textbf{Amazon}: This dataset comprises reviews, product descriptions, and images of various product categories \cite{mcauley2015image, he2016ups}. We consider three categories of products and obtain Beauty, Cell Phones and Clothing subsets. Each review rating is considered a positive user-item interaction. We preprocess the raw data by applying a 5-core filter on users and the statistics is presented in Table \ref{tab:dataset_statistics}. Visual information is represented using 4,096-dimensional features \cite{ni2019justifying}, and textual information is extracted using Sentence-Transformers \cite{reimers2019sentence}, resulting in 384-dimensional sentence embeddings. In each dataset, 20\% of the items are randomly chosen as cold-start items, which are further split into cold validation and testing sets in a 1:1 ratio. The remaining items are divided into training, warm validation, and warm testing sets with an 8:1:1 ratio \cite{wei2021contrastive}. KGs are constructed for each subset \cite{xian2019reinforcement} with 4 entity types and 6 relation types. The descriptions of each entity and relation are provided in Fig. \ref{fig:kg}. Feature entities from review data are preprocessed using TF-IDF to eliminate less meaningful words, retaining words with a frequency between 10 and 1,000 and a TF-IDF \cite{sparck1972statistical} score $>$ 0.1.

\textbf{Weixin-Sports:} The dataset consists of interaction data between users and sports-related micro-videos collected from Weixin Channels. User information is anonymized, retaining only anonymous IDs and relevant interaction samples. The dataset is split into training and evaluation sets, with all evaluation set users seen in the training set. Cold-start items in the evaluation set are separated into cold validation and testing sets in a 1:1 ratio, while a similar process is applied to warm-start items to create warm validation and testing sets. Preprocessing involves applying a 5-core filter to users and the statistics is presented in Table \ref{tab:dataset_statistics}. The dataset includes pre-extracted 64-dimensional multi-modal embeddings for the micro-videos. These micro-videos are linked to entities in a pre-built KG WikiSports through text matching between micro-video titles and entity names. The resulting KG for Weixin-Sports is constructed from the one-hop subgraph. Notably, WikiSports entities are closely related to sports, minimizing noisy knowledge and ensuring high-quality data.

\begin{table}[t]
\centering
\caption{Statistics of experimented datasets with constructed collaborative knowledge graphs.}
\label{tab:dataset_statistics}
\resizebox{1.0\columnwidth}{!}
{
\begin{tabular}{ccccc}
\hline \hline
Dataset                & Beauty   & Cell Phones & Clothing & Weixin-Sports\\  \hline
\#Users                & 22,363    & 27,879       & 39,387   & 336,466\\
\#Items                & 12,101    & 10,429       & 23,033   & 91,086 \\
\#Warm-start items           & 9,680     & 8,343        & 18,426   & 72,764 \\
\#Strict cold-start items           & 2,421     & 2,086        & 4,607    & 18,322 \\
\#Interactions         & 198,502   & 194,439      & 278,677  & 4,222,715 \\
\#Avg. Inter. of Users & 8.876    & 6.974       & 7.075   & 12.550 \\
\#Avg. Inter. of Items & 16.404   & 18.644      & 12.099  & 46.360 \\
Sparsity               & 99.927\% & 99.933\%    & 99.969\% & 99.986\%\\
\#Entities             & 748,114   & 702,194      & 3,012,459  & 1,612,249\\
\#Relations            & 7        & 7           & 7        & 227\\
\#Triplets              & 11,026,137 & 6,210,457     & 46,954,412 & 2,575,302\\ \hline \hline
\end{tabular}
}
\end{table}

\begin{table*}[t]
\centering
\caption{The strict cold-start and warm-start item recommendation performance comparison results on \textbf{Amazon} datasets. 'HM' represents the harmonic mean of metrics in two scenarios. The \textbf{bold} and the \underline{underline} show the best and second-best results within all comparison methods, respectively. All the numbers in the table are percentage numbers with '\%' omitted.}
\label{tab:main_results}
\resizebox{2.0\columnwidth}{!}{
\begin{tabular}{ccc|ccccc|ccccc|ccccc}
\hline \hline
\multirow{2}{*}{Setting} &
  \multirow{2}{*}{Type} &
  \multirow{2}{*}{Method} &
  \multicolumn{5}{c|}{Beauty} &
  \multicolumn{5}{c|}{Cell Phones} &
  \multicolumn{5}{c}{Clothing} \\
 &
   &
   &
  R@20 &
  M@20 &
  N@20 &
  H@20 &
  P@20 &
  R@20 &
  M@20 &
  N@20 &
  H@20 &
  P@20 &
  R@20 &
  M@20 &
  N@20 &
  H@20 &
  P@20 \\ \hline
\multirow{16}{*}{Cold} &
  \multirow{4}{*}{CF} &
  BPR \cite{rendle2009bpr} &
  0.94 &
  0.19 &
  0.33 &
  1.17 &
  0.06 &
  0.94 &
  0.21 &
  0.35 &
  1.11 &
  0.06 &
  0.42 &
  0.08 &
  0.15 &
  0.51 &
  0.03 \\
 &
   &
  LightGCN \cite{he2020lightgcn} &
  0.77 &
  0.14 &
  0.27 &
  0.97 &
  0.05 &
  1.12 &
  0.18 &
  0.37 &
  1.30 &
  0.07 &
  0.46 &
  0.10 &
  0.17 &
  0.55 &
  0.03 \\
 &
   &
  SGL \cite{wu2021self}&
  0.93 &
  0.21 &
  0.35 &
  1.12 &
  0.06 &
  0.79 &
  0.13 &
  0.26 &
  0.89 &
  0.04 &
  0.42 &
  0.11 &
  0.17 &
  0.51 &
  0.03 \\
 &
   &
  SimpleX \cite{mao2021simplex} &
  0.70 &
  0.15 &
  0.25 &
  0.87 &
  0.04 &
  1.01 &
  0.16 &
  0.33 &
  1.15 &
  0.06 &
  0.44 &
  0.08 &
  0.16 &
  0.51 &
  0.03 \\ \cline{2-18} 
 &
  \multirow{4}{*}{KG} &
  CKE \cite{zhang2016collaborative} &
  2.07 &
  0.48 &
  0.78 &
  2.73 &
  0.41 &
  1.58 &
  0.31 &
  0.55 &
  1.83 &
  0.09 &
  0.41 &
  0.11 &
  0.16 &
  0.51 &
  0.03 \\
 &
   &
  KGAT \cite{wang2019kgat}&
  {\ul 12.74} &
  {\ul 4.10} &
  {\ul 5.71} &
  {\ul 14.68} &
  {\ul 0.79} &
  {\ul 10.43} &
  {\ul 3.06} &
  {\ul 4.44} &
  {\ul 11.95} &
  {\ul 0.63} &
  {\ul 4.83} &
  {\ul 1.31} &
  {\ul 1.99} &
  {\ul 5.52} &
  {\ul 0.29} \\
 &
   &
  KGCN \cite{wang2019knowledge} &
  1.50 &
  0.36 &
  0.59 &
  1.79 &
  0.09 &
  1.95 &
  0.47 &
  0.74 &
  2.28 &
  0.11 &
  0.80 &
  0.22 &
  0.33 &
  0.98 &
  0.05 \\
 &
   &
  KGNNLS \cite{kgnnls}&
  1.71 &
  0.40 &
  0.66 &
  2.02 &
  0.10 &
  1.97 &
  0.47 &
  0.74 &
  2.29 &
  0.11 &
  0.80 &
  0.22 &
  0.33 &
  0.98 &
  0.05 \\ \cline{2-18} 
 &
  \multirow{4}{*}{MM} &
  VBPR \cite{he2016vbpr} &
  6.42 &
  2.57 &
  3.23 &
  7.76 &
  0.41 &
  4.33 &
  1.42 &
  1.98 &
  5.02 &
  0.25 &
  4.03 &
  1.34 &
  1.86 &
  4.67 &
  0.24 \\
 &
   &
  DRAGON \cite{zhou2023enhancing} &
  1.94 &
  0.54 &
  0.79 &
  2.30 &
  0.12 &
  0.83 &
  0.16 &
  0.30 &
  1.03 &
  0.05 &
  0.55 &
  0.11 &
  0.20 &
  0.65 &
  0.03 \\
 &
   &
  BM3 \cite{zhou2023bootstrap}&
  0.86 &
  0.19 &
  0.32 &
  1.02 &
  0.05 &
  1.28 &
  0.28 &
  0.49 &
  1.44 &
  0.07 &
  0.44 &
  0.10 &
  0.17 &
  0.58 &
  0.03 \\
 &
   &
  MMSSL \cite{wei2023multi} &
  0.67 &
  0.20 &
  0.28 &
  0.89 &
  0.04 &
  1.12 &
  0.20 &
  0.38 &
  1.32 &
  0.07 &
  0.52 &
  0.08 &
  0.17 &
  0.60 &
  0.03 \\ \cline{2-18} 
 &
  \multirow{2}{*}{CS} &
  DropoutNet \cite{volkovs2017dropoutnet} &
  4.85 &
  1.54 &
  2.15 &
  5.92 &
  0.31 &
  3.23 &
  0.68 &
  1.18 &
  3.65 &
  0.19 &
  2.15 &
  0.53 &
  0.86 &
  2.51 &
  0.13 \\
 &
   &
  CLCRec \cite{wei2021contrastive} &
  2.62 &
  0.59 &
  0.99 &
  3.16 &
  0.16 &
  1.97 &
  0.47 &
  0.77 &
  2.24 &
  0.11 &
  2.27 &
  0.65 &
  0.96 &
  2.66 &
  0.14 \\ \cline{2-18} 
 &
  \multirow{2}{*}{MM+KG} &
  MKGAT \cite{sun2020multi} &
  11.58 &
  3.59 &
  5.08 &
  13.46 &
  0.71 &
  4.61 &
  1.17 &
  1.84 &
  5.39 &
  0.03 &
  2.59 &
  0.74 &
  1.09 &
  3.11 &
  0.16 \\
 &
   &
  \textbf{Firzen (Ours)} &
  \textbf{13.65} &
  \textbf{5.00} &
  \textbf{6.54} &
  \textbf{15.83} &
  \textbf{0.85} &
  \textbf{11.97} &
  \textbf{3.38} &
  \textbf{5.08} &
  \textbf{13.44} &
  \textbf{0.70} &
  \textbf{8.15} &
  \textbf{2.63} &
  \textbf{3.64} &
  \textbf{9.39} &
  \textbf{0.49} \\ \hline
\multirow{16}{*}{Warm} &
  \multirow{4}{*}{CF} &
  BPR \cite{rendle2009bpr} &
  10.77 &
  4.87 &
  5.57 &
  14.70 &
  0.89 &
  12.34 &
  4.99 &
  6.14 &
  15.36 &
  0.83 &
  3.36 &
  1.38 &
  1.67 &
  4.34 &
  0.23 \\
 &
   &
  LightGCN \cite{he2020lightgcn}&
  13.20 &
  6.24 &
  6.96 &
  17.84 &
  1.08 &
  16.46 &
  6.71 &
  8.25 &
  20.28 &
  1.10 &
  5.88 &
  2.36 &
  2.88 &
  7.55 &
  0.40 \\
 &
   &
  SGL \cite{wu2021self}  &
  \textbf{14.56} &
  6.59 &
  {\ul 7.53} &
  \textbf{19.56} &
  1.18 &
  \textbf{16.91} &
  {\ul 6.92} &
  \textbf{8.51} &
  \textbf{20.79} &
  \textbf{1.12} &
  6.42 &
  2.56 &
  3.13 &
  8.34 &
  0.44 \\
 &
   &
  SimpleX \cite{mao2021simplex} &
  13.00 &
  5.42 &
  6.49 &
  17.54 &
  1.04 &
  15.73 &
  5.71 &
  7.41 &
  19.45 &
  1.05 &
  5.34 &
  1.79 &
  2.39 &
  6.89 &
  0.36 \\ \cline{2-18} 
 &
  \multirow{4}{*}{KG} &
  CKE \cite{zhang2016collaborative}&
  10.71 &
  4.93 &
  5.63 &
  14.63 &
  0.89 &
  12.00 &
  4.78 &
  5.94 &
  14.96 &
  0.81 &
  3.08 &
  1.24 &
  1.51 &
  3.98 &
  0.21 \\
 &
   &
  KGAT \cite{wang2019kgat} &
  11.29 &
  4.65 &
  5.53 &
  15.47 &
  0.92 &
  11.84 &
  4.19 &
  5.46 &
  14.88 &
  0.80 &
  4.06 &
  1.32 &
  1.79 &
  5.27 &
  0.27 \\
 &
   &
  KGCN \cite{wang2019knowledge} &
  8.88 &
  3.64 &
  4.35 &
  12.39 &
  0.75 &
  10.37 &
  3.66 &
  4.79 &
  13.08 &
  0.70 &
  2.25 &
  0.79 &
  1.03 &
  2.96 &
  0.15 \\
 &
   &
  KGNNLS \cite{kgnnls} &
  8.98 &
  3.63 &
  4.38 &
  12.48 &
  0.76 &
  10.37 &
  3.66 &
  4.79 &
  13.08 &
  0.70 &
  2.26 &
  0.79 &
  1.03 &
  2.96 &
  0.16 \\ \cline{2-18} 
 &
  \multirow{4}{*}{MM} &
  VBPR \cite{he2016vbpr} &
  8.95 &
  3.87 &
  4.51 &
  12.69 &
  0.76 &
  8.91 &
  3.16 &
  4.10 &
  11.31 &
  0.61 &
  2.84 &
  1.00 &
  1.28 &
  3.82 &
  0.20 \\
 &
   &
  DRAGON \cite{zhou2023enhancing} &
  9.39 &
  3.52 &
  4.41 &
  12.87 &
  0.74 &
  9.29 &
  2.94 &
  4.03 &
  11.81 &
  0.63 &
  4.11 &
  1.30 &
  1.78 &
  5.42 &
  0.28 \\
 &
   &
  BM3 \cite{zhou2023bootstrap} &
  13.49 &
  5.82 &
  6.81 &
  18.12 &
  1.05 &
  16.48 &
  6.61 &
  8.21 &
  20.16 &
  1.09 &
  {\ul 6.87} &
  {\ul 2.66} &
  {\ul 3.30} &
  {\ul 8.87} &
  {\ul 0.46} \\
 &
   &
  MMSSL \cite{wei2023multi} &
  14.30 &
  {\ul 6.60} &
  7.51 &
  19.28 &
  \textbf{1.20} &
  {\ul 16.82} &
  6.77 &
  8.34 &
  {\ul 20.76} &
  \textbf{1.12} &
  6.59 &
  2.58 &
  3.17 &
  8.56 &
  0.45 \\ \cline{2-18} 
 &
  \multirow{2}{*}{CS} &
  DropoutNet \cite{volkovs2017dropoutnet} &
  10.01 &
  4.07 &
  4.86 &
  14.00 &
  0.81 &
  12.38 &
  4.54 &
  5.80 &
  15.49 &
  0.82 &
  4.46 &
  1.61 &
  2.06 &
  5.87 &
  0.31 \\
 &
   &
  CLCRec \cite{wei2021contrastive} &
  3.39 &
  1.10 &
  1.47 &
  4.79 &
  0.25 &
  4.27 &
  1.26 &
  1.79 &
  5.46 &
  0.28 &
  0.82 &
  0.24 &
  0.34 &
  1.09 &
  0.06 \\ \cline{2-18} 
 &
  \multirow{2}{*}{MM+KG} &
  MKGAT \cite{sun2020multi}&
  11.08 &
  4.73 &
  5.54 &
  15.13 &
  0.90 &
  11.40 &
  3.92 &
  5.19 &
  14.35 &
  0.76 &
  3.66 &
  1.26 &
  1.65 &
  4.75 &
  0.24 \\
 &
   &
  \textbf{Firzen (Ours)} &
  {\ul 14.31} &
  \textbf{6.84} &
  \textbf{7.67} &
  {\ul 19.32} &
  {\ul 1.19} &
  16.75 &
  \textbf{6.93} &
  {\ul 8.44} &
  20.56 &
  {\ul 1.11} &
  \textbf{7.21} &
  \textbf{2.86} &
  \textbf{3.51} &
  \textbf{9.21} &
  \textbf{0.49} \\ \hline
\multirow{16}{*}{HM} &
  \multirow{4}{*}{CF} &
  BPR \cite{rendle2009bpr} &
  1.73 &
  0.37 &
  0.62 &
  2.17 &
  0.11 &
  1.75 &
  0.40 &
  0.66 &
  2.07 &
  0.11 &
  0.75 &
  0.15 &
  0.28 &
  0.91 &
  0.32 \\
 &
   &
  LightGCN \cite{he2020lightgcn}&
  1.46 &
  0.27 &
  0.52 &
  1.84 &
  0.10 &
  2.10 &
  0.35 &
  0.71 &
  2.44 &
  0.13 &
  0.85 &
  0.19 &
  0.32 &
  1.03 &
  0.06 \\
 &
   &
  SGL \cite{wu2021self}  &
  1.75 &
  0.41 &
  0.67 &
  2.12 &
  0.11 &
  1.51 &
  0.26 &
  0.50 &
  1.71 &
  0.08 &
  0.79 &
  0.21 &
  0.32 &
  0.96 &
  0.06 \\
 &
   &
  SimpleX \cite{mao2021simplex} &
  1.33 &
  0.29 &
  0.48 &
  1.66 &
  0.08 &
  1.90 &
  0.31 &
  0.63 &
  2.17 &
  0.11 &
  0.81 &
  0.15 &
  0.30 &
  0.95 &
  0.06 \\ \cline{2-18} 
 &
  \multirow{4}{*}{KG} &
  CKE \cite{zhang2016collaborative} &
  3.47 &
  0.87 &
  1.37 &
  4.60 &
  0.24 &
  2.79 &
  0.58 &
  1.01 &
  3.26 &
  0.16 &
  0.72 &
  0.20 &
  0.29 &
  0.90 &
  0.05 \\
 &
   &
  KGAT \cite{wang2019kgat} &
  {\ul 11.97} &
  {\ul 4.36} &
  {\ul 5.62} &
  {\ul 15.06} &
  {\ul 0.85} &
  {\ul 11.09} &
  {\ul 3.54} &
  {\ul 4.90} &
  {\ul 13.26} &
  {\ul 0.70} &
  {\ul 4.41} &
  {\ul 1.31} &
  {\ul 1.88} &
  {\ul 5.39} &
  {\ul 0.28} \\
 &
   &
  KGCN \cite{wang2019knowledge} &
  2.57 &
  0.66 &
  1.04 &
  3.13 &
  0.16 &
  3.28 &
  0.83 &
  1.28 &
  3.88 &
  0.19 &
  1.18 &
  0.34 &
  0.50 &
  1.47 &
  0.08 \\
 &
   &
  KGNNLS \cite{kgnnls} &
  2.87 &
  0.72 &
  1.15 &
  3.48 &
  0.18 &
  3.31 &
  0.83 &
  1.28 &
  3.90 &
  0.19 &
  1.18 &
  0.34 &
  0.50 &
  1.47 &
  0.08 \\ \cline{2-18} 
 &
  \multirow{4}{*}{MM} &
  VBPR \cite{he2016vbpr} &
  7.48 &
  3.09 &
  3.76 &
  9.63 &
  0.53 &
  5.83 &
  1.96 &
  2.67 &
  6.95 &
  0.35 &
  3.33 &
  1.15 &
  1.52 &
  4.20 &
  0.22 \\
 &
   &
  DRAGON \cite{zhou2023enhancing} &
  3.22 &
  0.94 &
  1.34 &
  3.90 &
  0.21 &
  1.52 &
  0.30 &
  0.56 &
  1.89 &
  0.09 &
  0.97 &
  0.20 &
  0.36 &
  1.16 &
  0.05 \\
 &
   &
  BM3 \cite{zhou2023bootstrap} &
  1.62 &
  0.37 &
  0.61 &
  1.93 &
  0.10 &
  2.38 &
  0.54 &
  0.92 &
  2.69 &
  0.13 &
  0.83 &
  0.19 &
  0.32 &
  1.09 &
  0.06 \\
 &
   &
  MMSSL \cite{wei2023multi} &
  1.28 &
  0.39 &
  0.54 &
  1.70 &
  0.08 &
  2.10 &
  0.39 &
  0.73 &
  2.48 &
  0.13 &
  0.96 &
  0.16 &
  0.32 &
  1.12 &
  0.06 \\ \cline{2-18} 
 &
  \multirow{2}{*}{CS} &
  DropoutNet \cite{volkovs2017dropoutnet} &
  6.53 &
  2.23 &
  2.98 &
  8.32 &
  0.45 &
  5.12 &
  1.18 &
  1.96 &
  5.91 &
  0.31 &
  2.90 &
  0.80 &
  1.21 &
  3.52 &
  0.18 \\
 &
   &
  CLCRec \cite{wei2021contrastive} &
  2.45 &
  0.88 &
  1.09 &
  3.83 &
  0.20 &
  2.70 &
  0.68 &
  1.08 &
  3.18 &
  0.16 &
  1.20 &
  0.35 &
  0.50 &
  1.55 &
  0.08 \\ \cline{2-18} 
 &
  \multirow{2}{*}{MM+KG} &
  MKGAT \cite{sun2020multi} &
  9.40 &
  4.20 &
  4.76 &
  13.48 &
  0.81 &
  6.57 &
  1.80 &
  2.72 &
  7.84 &
  0.05 &
  3.03 &
  0.93 &
  1.31 &
  3.76 &
  0.19 \\
 &
   &
  \textbf{Firzen (Ours)} &
  \textbf{13.97} &
  \textbf{5.78} &
  \textbf{7.06} &
  \textbf{17.40} &
  \textbf{0.99} &
  \textbf{13.96} &
  \textbf{4.54} &
  \textbf{6.34} &
  \textbf{16.25} &
  \textbf{0.86} &
  \textbf{7.65} &
  \textbf{2.74} &
  \textbf{3.57} &
  \textbf{9.30} &
  \textbf{0.49} \\ \hline \hline
\end{tabular}
}
\end{table*}

\textit{2) Setup and Evaluation Metrics:} We use the all-ranking protocol instead of the negative-sampling protocol to compute the evaluation metrics for recommendation performance comparison. In the warm-start recommendation setting, all warm-start items that have not been interacted by the given user are regarded as candidate items. In the cold-start recommendation setting, all cold-start items are regarded as candidate items. We use the commonly used metrics for Top-K recommendation performance Recall (R), Mean Reciprocal Ranking (M), Normalized Discounted Cumulative Gain (N), Hit Ratio (H) and Precision (P) at K=20. Moreover, we evaluate recommendation models with a metric aiming to balance performance between cold-start and warm-start using a harmonic mean of metrics in two settings, which equally weighs the importance of strict cold-start and warm-start recommendation, and penalizes models with a short barrel.

\textit{3) Base Models:} We consider five categories of recommendation models, including: (i) general CF recommendation models (CF): BPR \cite{rendle2009bpr}, LightGCN \cite{he2020lightgcn}, SGL \cite{wu2021self}, SimpleX \cite{mao2021simplex}, (ii) knowledge-aware recommendation models (KG): CKE \cite{zhang2016collaborative}, KGAT \cite{wang2019kgat}, KGCN \cite{wang2019knowledge}, KGNNLS \cite{kgnnls}, (iii) multi-modal recommendation models (MM): VBPR \cite{he2016vbpr}, DRAGON \cite{zhou2023enhancing}, BM3 \cite{zhou2023bootstrap}, MMSSL \cite{wei2023multi}, (iv) cold-start recommendation models: DropoutNet \cite{volkovs2017dropoutnet}, CLCRec \cite{wei2021contrastive} and (v) recommendation models based on both multi-modal content and KGs (MM+KG): MKGAT \cite{sun2020multi}. We implement all CS models based on LightGCN and use multi-modal content of items as side information for a fair comparison.

\subsection{Experimental Results: RQ1}

Table \ref{tab:main_results} and Table \ref{tab:Weixin_results} provide the performance comparison results of different methods on Amazon datasets and Weixin-Sports dataset, respectively.

\begin{table}[htbp]
\centering
\caption{The strict cold-start and warm-start item recommendation performance comparison results on \textbf{Weixin-Sports}.}
\label{tab:Weixin_results}
\resizebox{\columnwidth}{!}{
\begin{tabular}{ccc|ccccc}
\hline \hline
Setting                & Type                   & Method                 & R@20           & M@20           & N@20           & H@20           & P@20          \\ \hline
\multirow{16}{*}{Cold} & \multirow{4}{*}{CF}    & BPR \cite{rendle2009bpr}                    & 0.01           & 0.00           & 0.00           & 0.02           & 0.00          \\
                       &                        & LightGCN \cite{he2020lightgcn}              & 0.01           & 0.00           & 0.00           & 0.02           & 0.00          \\
                       &                        & SGL \cite{wu2021self}                    & 0.00           & 0.00           & 0.00           & 0.01           & 0.00          \\
                       &                        & SimpleX \cite{mao2021simplex}                & 0.00           & 0.00           & 0.00           & 0.00           & 0.00          \\ \cline{2-8} 
                       & \multirow{4}{*}{KG}    & CKE \cite{zhang2016collaborative}                    & 0.01           & 0.00           & 0.00           & 0.01           & 0.00          \\
                       &                        & KGAT \cite{wang2019kgat}                   & 0.03           & 0.01           & 0.01           & 0.04           & 0.00          \\
                       &                        & KGCN \cite{wang2019knowledge}                   & 0.01           & 0.00           & 0.00           & 0.02           & 0.00          \\
                       &                        & KGNNLS \cite{kgnnls}                 & 0.01           & 0.00           & 0.00           & 0.02           & 0.00          \\ \cline{2-8} 
                       & \multirow{4}{*}{MM}    & VBPR \cite{he2016vbpr}                   & {\ul 0.40}     & {\ul 0.11}     & {\ul 0.16}     & {\ul 0.06}     & {\ul 0.03}    \\
                       &                        & DRAGON \cite{zhou2023enhancing}                 & 0.08           & 0.04           & 0.04           & 0.15           & 0.01          \\
                       &                        & BM3 \cite{zhou2023bootstrap}                    & 0.02           & 0.00           & 0.01           & 0.02           & 0.00          \\
                       &                        & MMSSL \cite{wei2023multi}                  & 0.03           & 0.01           & 0.01           & 0.04           & 0.00          \\ \cline{2-8} 
                       & \multirow{2}{*}{CS}    & DropoutNet \cite{volkovs2017dropoutnet}             & 0.08           & 0.02           & 0.03           & 1.13           & 0.01          \\
                       &                        & CLCRec \cite{wei2021contrastive}                 & 0.05           & 0.01           & 0.02           & 0.08           & 0.00          \\ \cline{2-8} 
                       & \multirow{2}{*}{MM+KG} & MKGAT \cite{sun2020multi}                  & 0.01           & 0.00           & 0.00           & 0.01           & 0.00          \\
                       &                        & \textbf{Firzen (Ours)} & \textbf{0.48}  & \textbf{0.17}  & \textbf{0.21}  & \textbf{0.74}  & \textbf{0.04} \\ \hline
\multirow{16}{*}{Warm} & \multirow{4}{*}{CF}    & BPR \cite{rendle2009bpr}                   & 31.54          & 19.61          & 18.67          & 48.71          & 3.15          \\
                       &                        & LightGCN \cite{he2020lightgcn}               & 39.63          & 24.10          & 23.65          & 57.79          & 3.85          \\
                       &                        & SGL \cite{wu2021self}                    & 38.33          & 25.28          & 24.39          & 55.61          & 3.62          \\
                       &                        & SimpleX \cite{mao2021simplex}                & 23.73          & 13.30          & 13.22          & 38.25          & 2.32          \\ \cline{2-8} 
                       & \multirow{4}{*}{KG}    & CKE \cite{zhang2016collaborative}                    & 36.06          & 20.75          & 20.63          & 53.81          & 3.51          \\
                       &                        & KGAT \cite{wang2019kgat}                   & 36.82          & 22.78          & 22.19          & 54.61          & 3.57          \\
                       &                        & KGCN \cite{wang2019knowledge}                   & 30.73          & 16.23          & 16.61          & 46.47          & 2.90          \\
                       &                        & KGNNLS \cite{kgnnls}                 & 30.73          & 16.25          & 16.63          & 46.48          & 2.91          \\ \cline{2-8} 
                       & \multirow{4}{*}{MM}    & VBPR \cite{he2016vbpr}                   & 35.82          & 21.12          & 21.03          & 53.05          & 3.42          \\
                       &                        & DRAGON \cite{zhou2023enhancing}                 & 4.92           & 2.46           & 2.43           & 9.73           & 0.53          \\
                       &                        & BM3 \cite{zhou2023bootstrap}                    & 35.11          & 24.05          & 22.42          & 52.72          & 3.39          \\
                       &                        & MMSSL \cite{wei2023multi}                  & \textbf{50.72} & \textbf{36.75} & \textbf{34.80} & \textbf{69.10} & \textbf{5.01} \\ \cline{2-8} 
                       & \multirow{2}{*}{CS}    & DropoutNet \cite{volkovs2017dropoutnet}             & 37.79          & 22.83          & 22.29          & 55.88          & 3.68          \\
                       &                        & CLCRec \cite{wei2021contrastive}                 & 18.35          & 11.38          & 11.05          & 27.15          & 1.53          \\ \cline{2-8} 
                       & \multirow{2}{*}{MM+KG} & MKGAT \cite{sun2020multi}                  & 34.84          & 21.31          & 20.65          & 52.42          & 3.39          \\
                       &                        & \textbf{Firzen (Ours)} & {\ul 42.02}    & {\ul 30.12}    & {\ul 28.26}    & {\ul 60.38}    & {\ul 4.03}    \\ \hline
\multirow{16}{*}{HM}   & \multirow{4}{*}{CF}    & BPR \cite{rendle2009bpr}                    & 0.02           & 0.00           & 0.00           & 0.04           & 0.00          \\
                       &                        & LightGCN \cite{he2020lightgcn}              & 0.02           & 0.00           & 0.00           & 0.04           & 0.00          \\
                       &                        & SGL \cite{wu2021self}                    & 0.00           & 0.00           & 0.00           & 0.02           & 0.00          \\
                       &                        & SimpleX \cite{mao2021simplex}                & 0.00           & 0.00           & 0.00           & 0.00           & 0.00          \\ \cline{2-8} 
                       & \multirow{4}{*}{KG}    & CKE \cite{zhang2016collaborative}                    & 0.02           & 0.00           & 0.00           & 0.02           & 0.00          \\
                       &                        & KGAT \cite{wang2019kgat}                   & 0.06           & 0.02           & 0.02           & 0.08           & 0.00          \\
                       &                        & KGCN \cite{wang2019knowledge}                   & 0.02           & 0.00           & 0.00           & 0.04           & 0.00          \\
                       &                        & KGNNLS \cite{kgnnls}                 & 0.02           & 0.00           & 0.00           & 0.04           & 0.00          \\ \cline{2-8} 
                       & \multirow{4}{*}{MM}    & VBPR \cite{he2016vbpr}                   & {\ul 0.79}     & {\ul 0.22}     & {\ul 0.32}     & {\ul 0.12}     & {\ul 0.06}    \\
                       &                        & DRAGON \cite{zhou2023enhancing}                & 0.16           & 0.08           & 0.08           & 0.30           & 0.02          \\
                       &                        & BM3 \cite{zhou2023bootstrap}                    & 0.04           & 0.00           & 0.02           & 0.04           & 0.00          \\
                       &                        & MMSSL \cite{wei2023multi}                  & 0.06           & 0.02           & 0.02           & 0.08           & 0.00          \\ \cline{2-8} 
                       & \multirow{2}{*}{CS}    & DropoutNet \cite{volkovs2017dropoutnet}             & 0.16           & 0.04           & 0.06           & 0.26           & 0.02          \\
                       &                        & CLCRec \cite{wei2021contrastive}                 & 0.10           & 0.02           & 0.04           & 0.16           & 0.00          \\ \cline{2-8} 
                       & \multirow{2}{*}{MM+KG} & MKGAT \cite{sun2020multi}                  & 0.02           & 0.00           & 0.00           & 0.02           & 0.00          \\
                       &                        & \textbf{Firzen (Ours)} & \textbf{0.95}  & \textbf{0.34}  & \textbf{0.42}  & \textbf{1.46}  & \textbf{0.08} \\ \hline \hline
\end{tabular}
}
\end{table}

\textit{1) Main Results:} Firzen consistently improves strict cold-start scenarios and competes with state-of-the-art methods in warm-start scenarios. This validates the effectiveness of combining multiple sources of item information (visual, textual, and KGs) to balance performance. The diversity of evaluation datasets varies by recommendation domains, sparsity degrees and data scale, justifying the generality and flexibility of Firzen. The improvements are primarily due to two factors: i) SAGHL, enabling warm-start items to capture behavior-aware, modality-aware, and knowledge-aware collaborative signals, while strict cold-start items can capture interaction-related item-wise semantics through transformed multi-modal content and KG representation, ii) MSHGL, which transfers collaborative signals based on internal semantic structure, particularly from warm-start to strict cold-start items.

\textit{2) Performance Comparison with KG and MM Models:} Compared to CF methods, most knowledge-aware and multi-modal models outperform CF in either warm-start or strict cold-start scenarios, demonstrating the value of incorporating multi-modal content and knowledge graphs to address sparsity. KGAT excels in strict cold-start recommendations by using external semantic knowledge to encode item-wise structures, while MMSSL performs well in warm-start scenarios by modeling modality-specific user preferences and item characteristics. However, KGAT underperforms in warm-start situations due to potential unrelated external knowledge, and MMSSL struggles with strict cold-start item recommendations, as it relies on a complete user-item interaction graph, making it challenging to generate reasonable embeddings for new items.

\textit{3) Performance Comparison with CS Models:} Table \ref{tab:main_results} shows that cold-start models incorporating additional content-based representations bring a significant improvement in strict cold-start but hurt the warm-start ones compared to the backbone LightGCN. The reason is that these models attempt to produce compromise representations for strict cold-start with no interaction records and warm-start items with sufficient ones. Instead of representing strict cold-start and warm-start items with behavioral and content features  respectively, our Firzen bridges the gap via extracting interaction-related content features to enhance behavior representations of warm-start items, while propagates behavioral features to enhance content representations of strict cold-start ones.

\textit{4) Performance Comparison with MM+KG Models:} Integrating multi-modal content represented as nodes into KGs, MKGAT suffers from the performance degradation, as the number of multi-modal content is much lower than that of entities in KG, making it difficult to propagate sufficient and effective information to item and user nodes. Different from MKGAT, our Firzen effectively collaborates the multi-modal content and KGs from two aspects: i) The interaction-related content including multi-modal content and knowledge-aware item representation are independently extracted and fused in an importance-aware manner; ii) The modality-aware latent structure is mined to transfer the interaction-related content including knowledge-aware item representation. 

\begin{table}[t]
\centering
\caption{Ablation study of different components.}
\label{tab:ablation}
\resizebox{1.0\columnwidth}{!}
{
\begin{tabular}{cccccccccc}
\hline \hline
BA                 & KA                 & MA                 & MS                 & Setting & R@20  & M@20 & N@20 & H@20  & P@20 \\ \hline
\multirow{3}{*}{\checkmark} & \multirow{3}{*}{\checkmark} & \multirow{3}{*}{\checkmark} & \multirow{3}{*}{} & Cold    & 3.31     & 0.86    & 1.35    & 3.93     & 0.20    \\
                   &                    &                    &                    & Warm    & 14.27     & 6.72    & 7.59    & 19.32     & 1.18    \\
                   &                    &                    &                    & HM      & 5.37     & 1.52    & 2.29    & 6.53     & 0.34    \\ \hline
\multirow{3}{*}{\checkmark} & \multirow{3}{*}{\checkmark} & \multirow{3}{*}{} & \multirow{3}{*}{\checkmark} & Cold    & 12.93     & 4.77    & 6.23    & 14.93     & 0.80    \\
                   &                    &                    &                    & Warm    & 14.29     & 6.93    & 7.71    & 19.21     & 1.18    \\
                   &                    &                    &                    & HM      & 13.58     & 5.65    & 6.89    & 16.80     & 0.95    \\ \hline
\multirow{3}{*}{\checkmark} & \multirow{3}{*}{} & \multirow{3}{*}{\checkmark} & \multirow{3}{*}{\checkmark} & Cold    & 11.63     & 4.36    & 5.63    & 13.61     & 0.73    \\
                   &                    &                    &                    & Warm    & 14.38     & 6.75    & 7.62    & 19.42     & 1.18    \\
                   &                    &                    &                    & HM      & 12.86     & 5.30    & 6.48    & 16.00     & 0.90    \\ \hline
\multirow{3}{*}{} & \multirow{3}{*}{\checkmark} & \multirow{3}{*}{\checkmark} & \multirow{3}{*}{\checkmark} & Cold    & 12.35     & 4.37    & 5.82    & 14.29     & 0.76    \\
                   &                    &                    &                    & Warm    & 13.13     & 5.49    & 6.50    & 17.91     & 1.07    \\
                   &                    &                    &                    & HM      & 12.73     & 4.87    & 6.14    & 15.90     & 0.89    \\ \hline
\multirow{3}{*}{\checkmark} & \multirow{3}{*}{\checkmark} & \multirow{3}{*}{\checkmark} & \multirow{3}{*}{\checkmark} & Cold & 13.65 & 5.00 & 6.54 & 15.83 & 0.85 \\
                   &                    &                    &                    & Warm    & 14.31 & 6.84 & 7.67 & 19.32 & 1.19 \\
                   &                    &                    &                    & HM      & \textbf{13.97} & \textbf{5.78} & \textbf{7.06} & \textbf{17.40} & \textbf{0.99} \\ \hline \hline
\end{tabular}
}
\end{table}

\subsection{Ablation Study: RQ2}

We perform an ablation study on the dataset Beauty to evaluate the effectiveness of different components in Firzen on the recommendation performance. Specifically, we consider the following variants: (1) \textbf{w/o MS}: Removing MSHGL, (2) \textbf{w/o MA}: Removing modality-aware graph convolution, (3) \textbf{w/o KA}: Removing knowledge-aware graph attention, and (4) \textbf{w/o BA}: Removing behavior-aware graph convolution.

As shown in Table \ref{tab:ablation}, Firzen performs best among all the comparison methods on HM of metrics in strict cold-start and warm-start scenarios, which shows that the removal of any components from Firzen will hurt the final performance. The effectiveness of MS is significant, which reflects that the information propagation and aggregation from warm-start items to cold-start items is the foundation for warming up the latter. Removing MA and KA both decrease the performance on cold-start scenario, but have no significant influence on warm-start scenario. This reflects that multi-modal content and KGs have complementary information to characterize the items in a relatively comprehensive manner. We also find that Firzen without BA performs worse both on cold-start and warm-start scenarios. The above two observations show that items' side information can be beneficial for cold-start recommendation, while that could also impair the quality of item representation from two aspects. On the hand, directly utilizing multi-modal content or KG representations will inevitably bring noise in preserving key semantic of items related to the recommendation task. On the other hand, the information fusion of multi-modal content and KG content is easily towards over-smoothed, since the overwhelming information can make the collaborative embeddings of items blurred\cite{zhou2023contrastive}. For instance, two movies share the visually similar posters but the user only prefers one of them as he appreciates the film's director, then the final representations of two items are blurrier. Our method is less sensitive to the interaction-unrelated content as behavior-aware user and item representations are incorporated to model the collaborative signals directly from the interactions.

\subsection{Parameter Sensitivity: RQ3}

We investigate the sensitivity of the fusing weight of knowledge-aware user/item representations $\lambda_{k}$, that of modality-aware user/item representations $\lambda_{m}$, and the momentum to control the update of modality importance $\eta$ in SAHGL on Firzen performance. Moreover, different number of neighbors per item $K$ for MSHGL is investigated. 
\begin{figure}[t]
\centering
{
    \subfigure[$\lambda_{k}$]{
       \centering
        \includegraphics[width=0.4\linewidth]{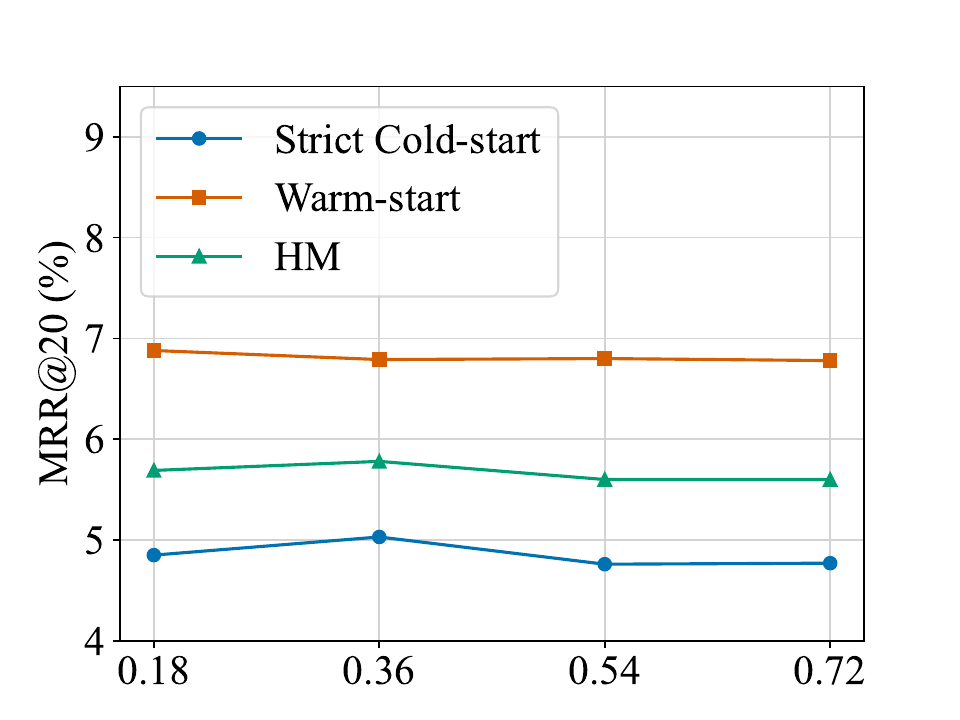}
    }
    \subfigure[$\lambda_{m}$]{
   \centering
    \includegraphics[width=0.4\linewidth]{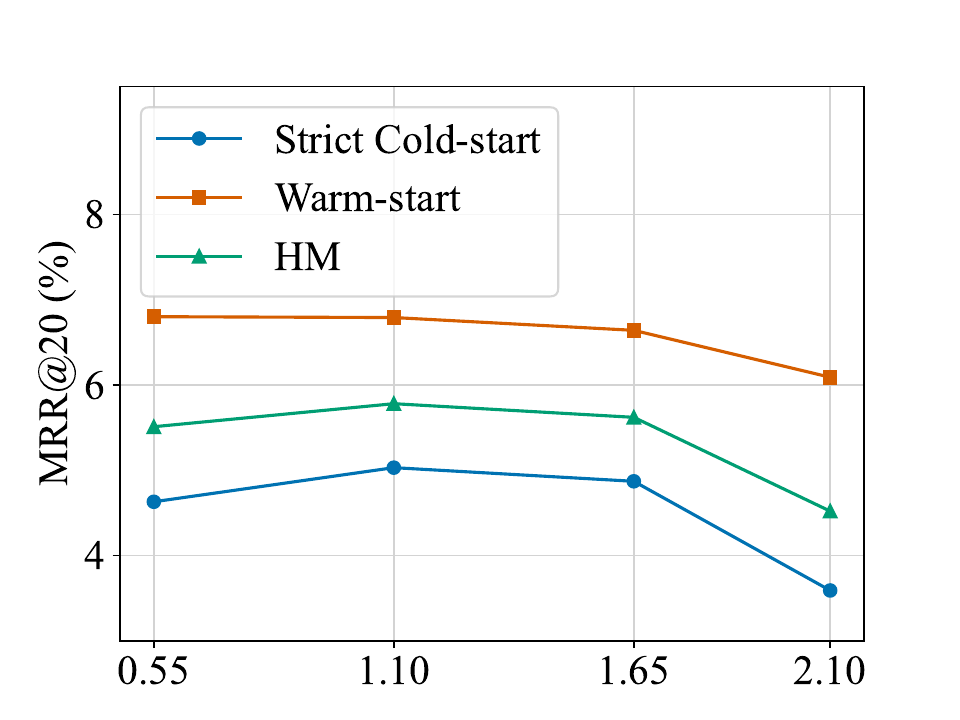}
    }    \\
   \vspace{-0.36cm}
    \subfigure[$\eta$]{
   \centering
    \includegraphics[width=0.4\linewidth]{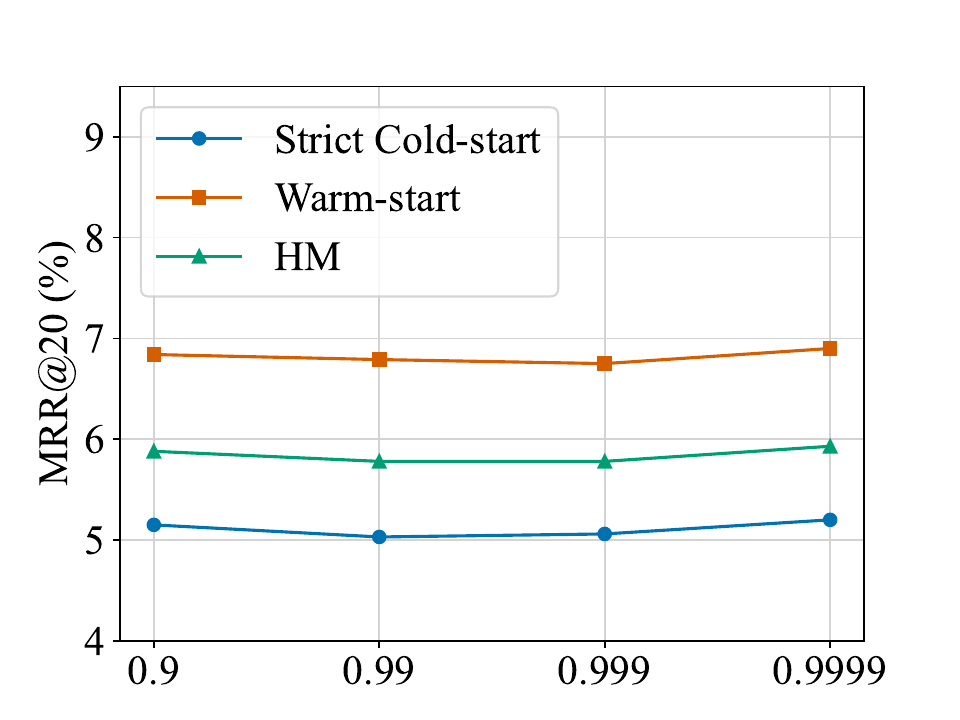}
    }
    \subfigure[$K$]{
   \centering
    \includegraphics[width=0.4\linewidth]{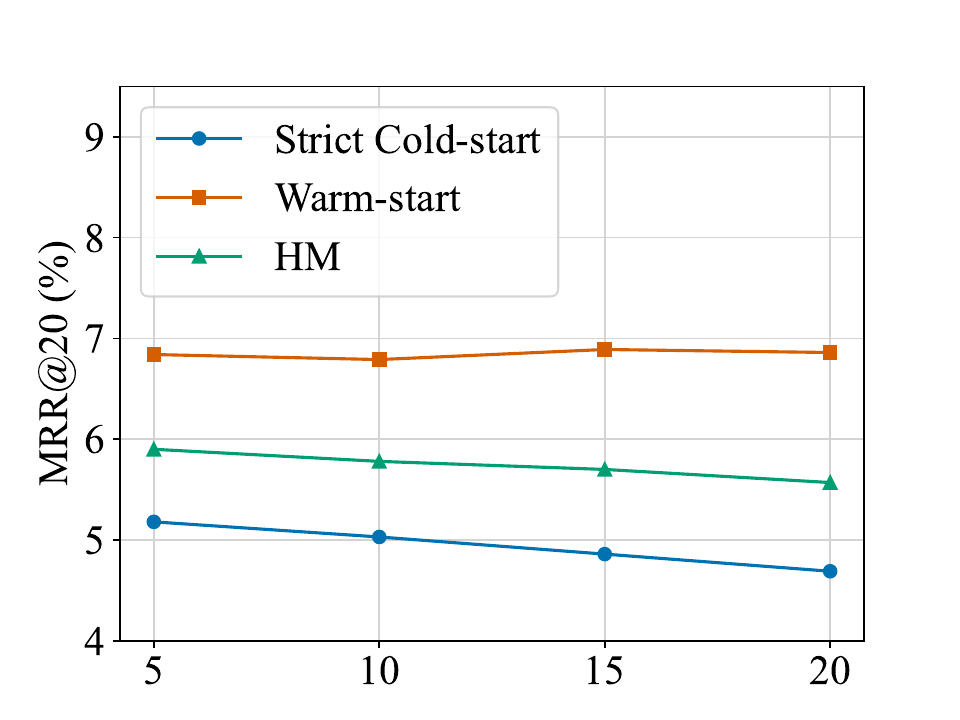}
    }

}\caption{Performance with different hyperparameters.}
\label{fig:hyperparameter}
\end{figure}

\textit{1) Hyperparameters $\lambda_{k}$ and $\lambda_{m}$}: We first vary the value of $\lambda_{k}$ in \{0.18, 0.36, 0.54. 0.72\} with $\lambda_{m}$ set to 1.10, then vary the value of $\lambda_{m}$ in $\{0.55, 1.10, 1.65, 2.20\}$ with $\lambda_{k}$ set to 0.36. The curves of $\lambda_{k}$ and $\lambda_{m}$ are shown in Fig. \ref{fig:hyperparameter} (a) and (b), respectively, from which we observe that the performance in strict cold-start scenario increases first with the increasing of $\lambda_{k}$ and $\lambda_{m}$, and decreases when $\lambda_{k}$ is over 0.36 or $\lambda_{m}$ is over 1.10, showing the benefit of fusing side information and collaborative signals in a proper ratio. However, the performance in warm-start scenario decreases with the increasing of $\lambda_{k}$ and $\lambda_{m}$, as unrelated multi-modal content or entities in KGs may blur the final representations.

\textit{2) Momentum $\eta$ and Neighbor Number $K$:} We vary the value of $\eta$ in \{0.9, 0.99, 0.999, 0.9999\} while the value of $K$ is chosen from \{5, 10, 15, 20\}. The curves of $\eta$ and $K$ are shown in Fig. \ref{fig:hyperparameter} (c) and (d), respectively. We can see that the model performance is insensitive to changes in $\eta$. We observe the strict cold-start performance degradation with the increase of $K$, which shows the over-connection of homogeneous graph could result in improper information propagation from warm-start to strict cold-start items. Representations of strict cold-start items are sensitive to the potential association with warm-start neighbors, and thus even small noise or relatively improper connection can have a serious negative impact on their final representations.

\subsection{Robustness Analysis: RQ4}
We conduct experiments to illustrate the proposed model’s robustness to outliers, duplicates or discrepancies in entities within the KG. Specifically, we inject 20\% noisy triplets to the constructed KG in three forms, respectively: (1) outliers: triplets with non-existent tail entities (e.g., new brands or categories), (2) duplicates: triplets same with existing ones, (3) discrepancies: triplets with existing but invalid tail entities (e.g., incorrect brands or categories). The experimental results on the noisy KG are reported in Table \ref{tab:noise}. Despite the KG noise, Firzen still \textbf{achieves the best performance} on M@20 when competing with state-of-the-art knowledge-aware recommendation models, in distilling the interaction-related information from noisy KG to assist the modeling of user preference and item characteristics. Additionally, Firzen \textbf{achieves lowest average performance decreasing in alleviating KG noise}, which verifies the superiority in discovering relevant and useful item semantics from noisy KG information.

\begin{table}[t]
\centering
\caption{Performance in alleviating KG noise. 'Avg. Dec.' represents the performance degradation percentage in terms of M@20.}
\label{tab:noise}
\resizebox{1.0\columnwidth}{!}
{
\begin{tabular}{ccc|cccccc}
\hline \hline
\multirow{2}{*}{Setting} & \multirow{2}{*}{Type} & \multirow{2}{*}{Method} & \multicolumn{2}{c}{Outlier} & \multicolumn{2}{c}{Duplicate} & \multicolumn{2}{c}{Discrepancy} \\ \cline{4-9} 
                      &                        &                        & M@20          & Avg.Dec $\downarrow$        & M@20          & Avg.Dec $\downarrow$              & M@20       & Avg.Dec $\downarrow$ \\ \hline
\multirow{6}{*}{Cold} & \multirow{4}{*}{KG}    & CKE \cite{zhang2016collaborative}                    & 0.24    & 50.00          &  0.25    &  47.92          & 0.30       & 37.50   \\
                      &                        & KGAT \cite{wang2019kgat}                  & 0.14          & 96.59          & 0.18          & 95.61                & 0.40       & 90.24   \\
                      &                        & KGCN \cite{wang2019knowledge}                   & 0.22          & 38.89 & 0.13          & 63.89                & 0.15       & 58.33   \\
                      &                        & KGNNLS \cite{kgnnls}                 & 0.22          & 45.00          & 0.13          & 67.50                & 0.15       & 62.50   \\ \cline{2-9} 
                      & \multirow{2}{*}{MM+KG} & MKGAT \cite{sun2020multi}                  & 0.24    & 93.31          & 0.22          & 93.87                & 0.24       & 93.31   \\
                      &                        & \textbf{Firzen (Ours)} & 2.81 & 43.80    & 2.86 & 42.80        & 2.69  & 46.20         \\ \hline
\multirow{6}{*}{Warm} & \multirow{4}{*}{KG}    & CKE \cite{zhang2016collaborative}                    & 4.65          & 5.68           & 4.81          & 2.43                 & 4.73       & 4.06    \\
                      &                        & KGAT \cite{wang2019kgat}                   & 3.72          & 20.00          & 3.77          & 18.92                & 4.09       & 12.04   \\
                      &                        & KGCN \cite{wang2019knowledge}                   & 3.63          & 0.27     & 3.64          & 0.00           & 3.50       & 3.58    \\
                      &                        & KGNNLS \cite{kgnnls}                 & 3.63          & 0.00  & 3.64          & -0.28 & 3.50       & 3.58    \\ \cline{2-9} 
                      & \multirow{2}{*}{MM+KG} & MKGAT \cite{sun2020multi}                  & 3.69          & 21.99          & 4.33          & 8.46                 & 3.99       & 15.64   \\
                      &                        & \textbf{Firzen (Ours)} & 6.28 & 8.19           &6.38 & 6.73                & 6.23     & 8.92     \\ \hline
\multirow{6}{*}{HM}      & \multirow{4}{*}{KG}   & CKE \cite{zhang2016collaborative}                     & {\ul 0.46}      & 47.13     & {\ul 0.48}    & {\ul 51.72}   & 0.56        & {\ul 48.28}       \\
                      &                        & KGAT \cite{wang2019kgat}                   & 0.27          & 93.81          & 0.34          & 92.20                & {\ul 0.73} & 83.26   \\
                      &                        & KGCN \cite{wang2019knowledge}                   & 0.41          & {\ul 37.88}    & 0.25          & 62.12                & 0.29       & 56.06   \\
                      &                        & KGNNLS \cite{kgnnls}                 & 0.41          & 43.06          & 0.25          & 65.28                & 0.29       & 59.72   \\ \cline{2-9} 
                      & \multirow{2}{*}{MM+KG} & MKGAT \cite{sun2020multi}                  & 0.45          & 89.29          & 0.42          & 90.00                & 0.45       & 89.29   \\
                      &                        & \textbf{Firzen (Ours)} & \textbf{3.88} & \textbf{32.87} & \textbf{3.95} & \textbf{31.66}       & \textbf{3.76}  & \textbf{34.95}         \\ \hline \hline
\end{tabular}
}
\end{table}

\begin{table}[t]
\centering
\caption{The normal cold-start item recommendation performance.}
\label{tab:normal_coldstart}
\resizebox{\columnwidth}{!}{
\begin{tabular}{cc|ccccc}
\hline \hline
Type                   & Method                 & R@20           & M@20           & N@20           & H@20           & P@20          \\ \hline
\multirow{4}{*}{CF}    & BPR \cite{rendle2009bpr}                   &  1.03          &  0.20          &   0.38     & 1.05           & 0.05          \\
                    & LightGCN \cite{he2020lightgcn}              &  10.35          & \uline{3.93}           & \uline{5.24}           & 11.21           & 0.58          \\
                                             & SGL \cite{wu2021self}                    &   9.68       & 3.47           & 4.77           & 10.36           & 0.52          \\
                                           & SimpleX \cite{mao2021simplex}                &  0.70          & 0.15           & 0.25           & 0.87           & 0.04          \\ \cline{1-7} 
                      \multirow{4}{*}{KG}    & CKE \cite{zhang2016collaborative}                    &  2.19        & 0.33           & 0.71           & 2.48           & 0.12          \\
                                            & KGAT \cite{wang2019kgat}                   & 10.85          & 3.31           & 4.81           & 11.71           & 0.60          \\
                                      & KGCN \cite{wang2019knowledge}                   &  1.47         & 0.33           & 0.56           & 1.66           & 0.08          \\
                                     & KGNNLS \cite{kgnnls}                 &  1.53         & 0.36           & 0.60           & 1.70           & 0.08          \\ \cline{1-7} 
                   \multirow{4}{*}{MM}    & VBPR \cite{he2016vbpr}                   &  6.19   & 2.36     & 3.08     & 6.79     & 0.34    \\
                                        & DRAGON \cite{zhou2023enhancing}                 &  8.02        & 2.19           & 3.33           & 8.79           & 0.45          \\
                                           & BM3 \cite{zhou2023bootstrap}                    &  6.77        & 2.14           & 3.10           & 7.27           & 0.37          \\
                                            & MMSSL \cite{wei2023multi}                  &   9.73        & 3.46           & 4.78           & 10.46           & 0.54          \\ \cline{1-7} 
                       \multirow{2}{*}{CS}    & DropoutNet \cite{volkovs2017dropoutnet}             &  4.62       & 1.16           & 1.84           & 5.16           & 0.26          \\
                                            & CLCRec \cite{wei2021contrastive}                 &  2.21        & 0.56           & 0.89           & 2.41           & 0.12          \\ \cline{1-7} 
                      \multirow{2}{*}{MM+KG} & MKGAT \cite{sun2020multi}                  &  \uline{11.19}         & 3.39           & 4.98           & \uline{12.09}           & \uline{0.62}          \\
                                             & \textbf{Firzen (Ours)} & \textbf{13.40}  & \textbf{4.70}  & \textbf{6.45}  & \textbf{14.36}  & \textbf{0.74} \\ \hline \hline

\end{tabular}
}
\end{table}

\subsection{Generalizability Analysis: RQ5}
Though Firzen focuses on warming up strict cold-start items, it can be seamlessly utilized to enhance the normal cold-start item recommendation with additional user-item link provided at the test stage. We conduct experiments to illustrate the effectiveness of Firzen in the normal cold-start item recommendation, shown as Table  \ref{tab:normal_coldstart}. Specifically, we further split the cold validation and testing sets into \textit{known} and \textit{unknown} sets with the ratio of 1:1, respectively. The known set simulates the newly added interaction data of normal cold-start items that can be utilized at the inference stage, while the unknown set is used for performance comparison. We observe that some recommendation methods incorporating the user-item interaction graph (e.g., LightGCN, MMSSL) would obtain performance gain compared with strict cold-start scenario. Due to the information propagation between users and cold-start items, Firzen can still \textbf{achieve state-of-the-art performance in normal cold-start scenario}, demonstrating the effectiveness in balancing the recommendation performance of warm-start and cold-start (both strict and normal).

\subsection{Training and Inference Time: RQ6}
We show the training time and inference time per user measured on 1 TITAN Xp GPU to investigate how they are affected when integrating multiple sources of additional information, as shown in Table \ref{tab:time}.

Since KG representation, adversarial and contrastive loss are introduced when integrating knowledge-aware modality (KA), the training time relatively increases compared with using mere behavior-aware collaborative signals (BA). Due to the knowledge graph attention, the inference time for cold-start increases for KA, while the modality-aware graph convolution along with MSHGL introduced by visual modalities (VA) and textual modality (TA) \textbf{bring insignificant inference latency, which satisfies most real-time requirements.}

\begin{table}[t]
\centering
\caption{Training and Inference Time on Amazon Beauty.}
\label{tab:time}
\resizebox{1.0\columnwidth}{!}
{
\begin{tabular}{cccc|ccc}
\hline \hline
\multirow{2}{*}{BA} & \multirow{2}{*}{KA} & \multirow{2}{*}{VA} & \multirow{2}{*}{TA} & \multirow{2}{*}{Training Time (s)} & \multicolumn{2}{c}{Inference Time Per User (ms)} \\ \cline{6-7}
 &  &  &  &          & Cold & Warm \\ \hline 
 \checkmark  &  &  &  & 8124.24  & 11.41     & 51.66     \\
 \checkmark  & \checkmark   &  &  & 29409.30 & 15.77 & 52.76     \\
 \checkmark  & \checkmark &  \checkmark  &  & 44497.11 & 16.04     & 57.56     \\
 \checkmark  &\checkmark  & \checkmark &  \checkmark  & 48271.68 & 16.27     & 57.85     \\ \hline \hline
\end{tabular}
}
\end{table}

\begin{figure}[t]
  \centering
  \includegraphics[width=0.9\linewidth]{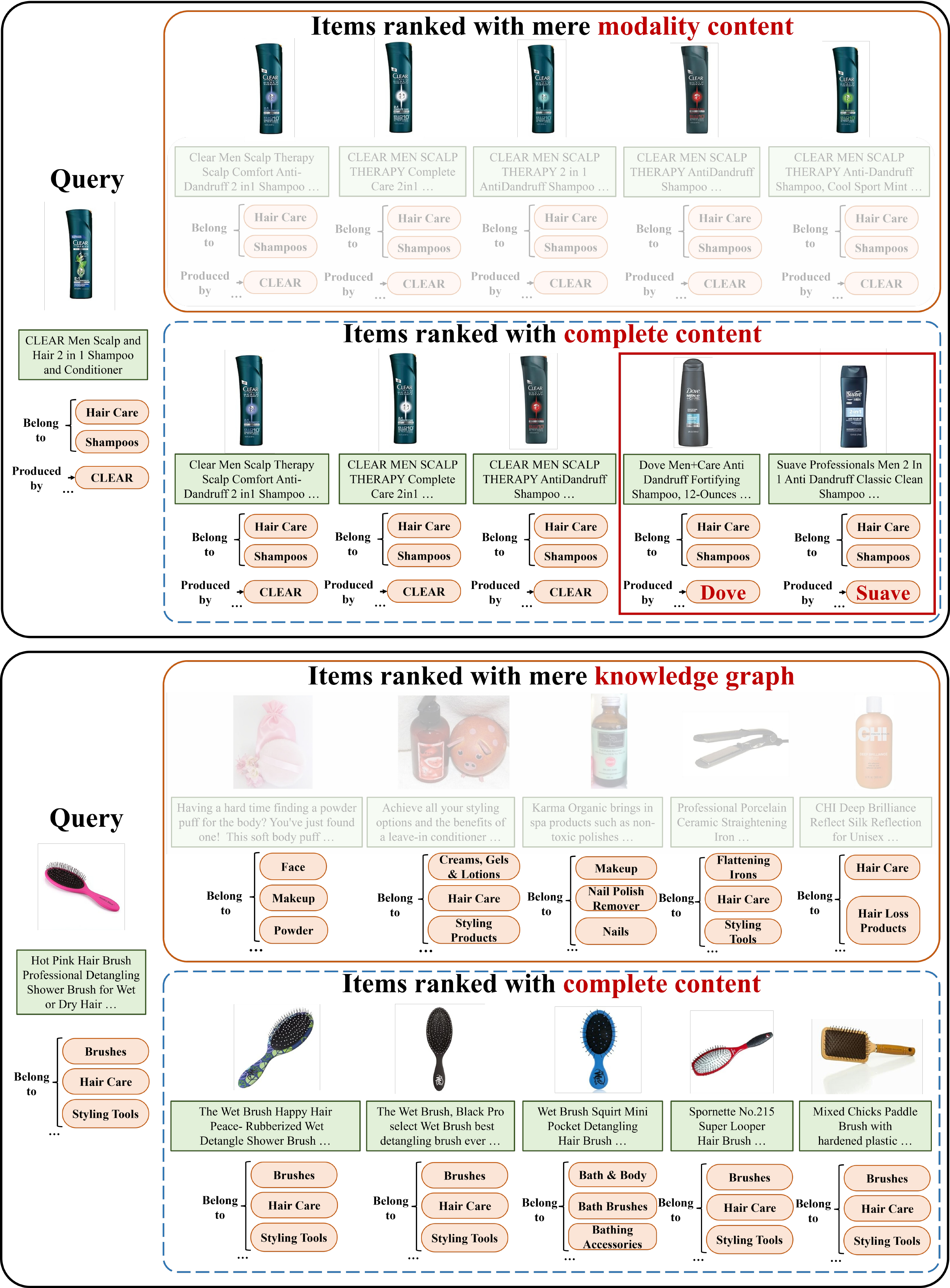}
   \caption{Examples of relevant items ranked with mere modality content, KG or complete content. The transparency indicates unused side information.}
   \label{fig:case_all}
\end{figure}

\subsection{Interpretability: RQ7}
We perform qualitative case study and quantitative ablation study to explain the interpretability. \textit{\textbf{(1) Case study:}} We first illustrate five most similar items with the sampled item from Amazon Beauty, to explain how the model would provide unexpected or unsatisfied suggestions without complete content, as illustrated in Figure \ref{fig:case_all}.
We observe that the modality content and KG can collaborate to \textbf{balance the diversity and relevance} of recommendation. The first case shows that with mere modality, the five most similar items are all shampoos of the same brand. Nevertheless, taking textual content and KG into consideration, interaction-related textual content and external knowledge are extracted by modality-aware graph convolution and knowledge graph attention to recommend shampoos \textbf{with more diversified brands} since they all belong to Hair Care and Shampoos. The second case shows that with mere knowledge graph, some items irrelevant to the brush rank first due to the inevitable KG noise. However, the visual and textual content are utilized in modality-aware graph convolution and modality-specific graph convolution to constrain the relevant brushes \textbf{with different appearances} to be ranked first. \textbf{(2) Ablation study:} We conduct experiments to interpret the contribution of different modalities and KG information to the final recommendations. Specifically, we inference by gradually incorporating items' textual features, visual features or knowledge-aware features. We observe that \textbf{different modalities and KG information all contribute to the performance gain}. Moreover, different side information differs in the amount of distribution. For instance, based on the behavior-aware collaborative signals (BA), the textual modality (TA) contributes more than visual modality (VA) and knowledge-aware modality (KA) on Amazon Beauty, since the items seem similar in visual content or densely connected in KG, but are more distinctive in textual descriptions.

\begin{table}[t]
\centering
\caption{Contribution of different modality and KG information.}
\label{tab:contribution}
\resizebox{1.0\columnwidth}{!}
{
\begin{tabular}{cccccccccc}
\hline \hline
BA                 & KA                 & VA        & TA                          & Setting & R@20  & M@20 & N@20 & H@20  & P@20 \\ \hline
                    \multirow{3}{*}{\checkmark} & \multirow{3}{*}{} & \multirow{3}{*}{}  & \multirow{3}{*}{} &  Cold    & 0.36     & 0.05    & 0.11    & 0.48     & 0.02    \\
                   &                    &           &                       & Warm    & 12.81     & 5.76    & 6.55    & 17.63     & 1.07    \\
                   &                    &           &                        & HM      & 0.70     & 0.10    & 0.22    & 0.93     & 0.04    \\ \hline
                   \multirow{3}{*}{\checkmark} & \multirow{3}{*}{\checkmark} & \multirow{3}{*}{}  & \multirow{3}{*}{} &  Cold    & 1.70     & 0.41    & 0.66    & 2.10     & 0.11    \\
                   &                    &           &                        & Warm    & 13.20     & 6.22    & 6.95    & 18.13     & 1.10    \\
                   &                    &           &                         & HM      & 3.01     & 0.77    & 1.21    & 3.76     & 0.20    \\ \hline
                   \multirow{3}{*}{\checkmark} & \multirow{3}{*}{} & \multirow{3}{*}{\checkmark}  & \multirow{3}{*}{} &  Cold    & 7.84     & 2.55    & 3.49    & 9.39     & 0.50    \\
                   &                    &           &                        & Warm    & 12.90     & 5.93    & 6.70    & 17.58     & 1.05    \\
                   &                    &           &                         & HM      &  9.75    & 3.57    & 4.59    & 12.24     & 0.68    \\ \hline
                   \multirow{3}{*}{\checkmark} & \multirow{3}{*}{} & \multirow{3}{*}{}  & \multirow{3}{*}{\checkmark} &  Cold    & 12.04     & 4.22    & 5.64    & 14.02     & 0.75    \\
                   &                    &           &                         & Warm    & 13.55     & 6.30    & 7.10    & 18.31     & 1.10    \\
                   &                    &           &                        & HM      &  12.75    & 5.05    & 6.29    & 15.88     & 0.89    \\ \hline \hline
\end{tabular}
}
\end{table}

\subsection{Visualization: RQ8}
\begin{figure}[t]
\vspace{-0.22cm}
\centering
{   \centering
    \subfigure[LightGCN]{
       \centering
        \includegraphics[width=0.25\linewidth]{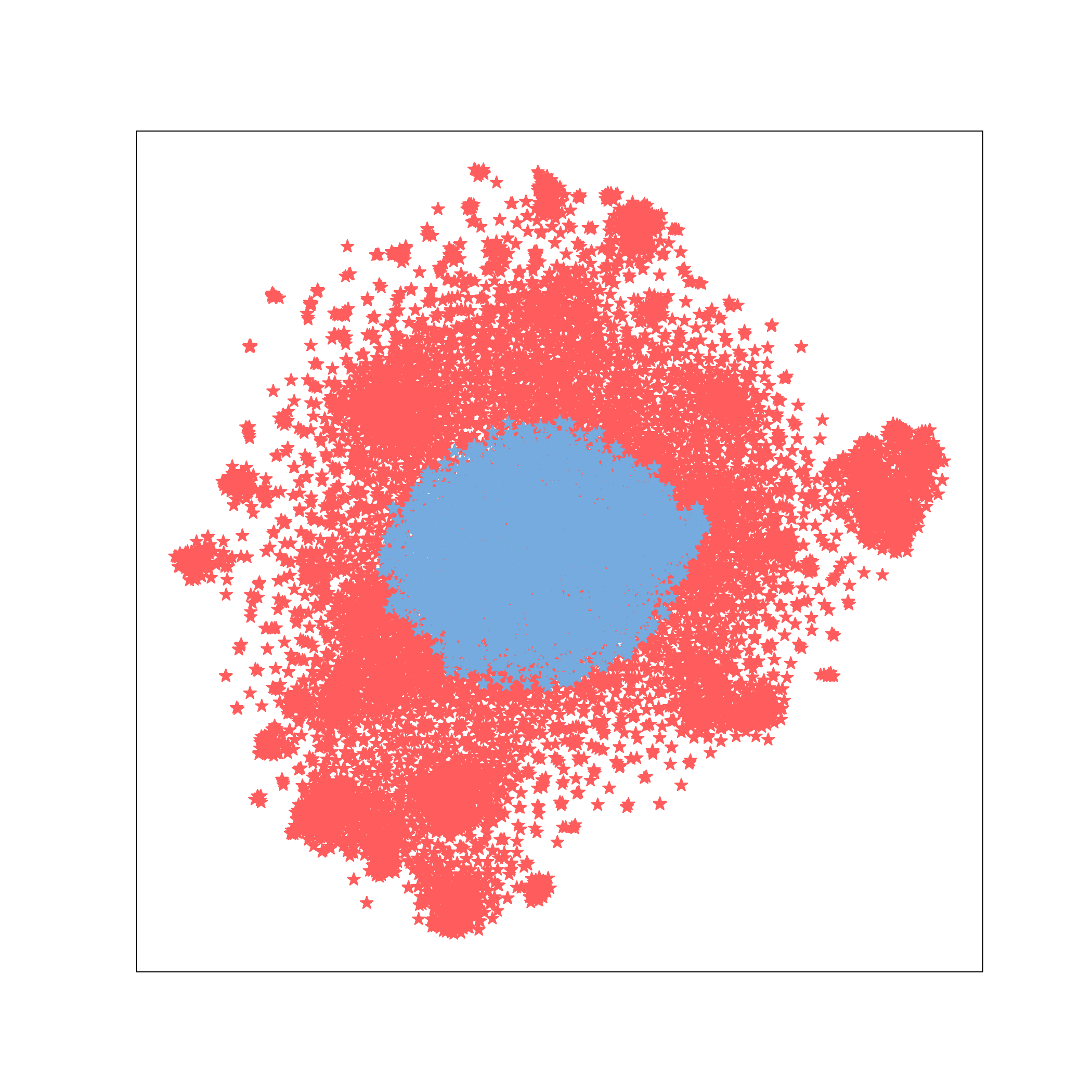}
    }
    \subfigure[KGAT]{
   \centering
    \includegraphics[width=0.25\linewidth]{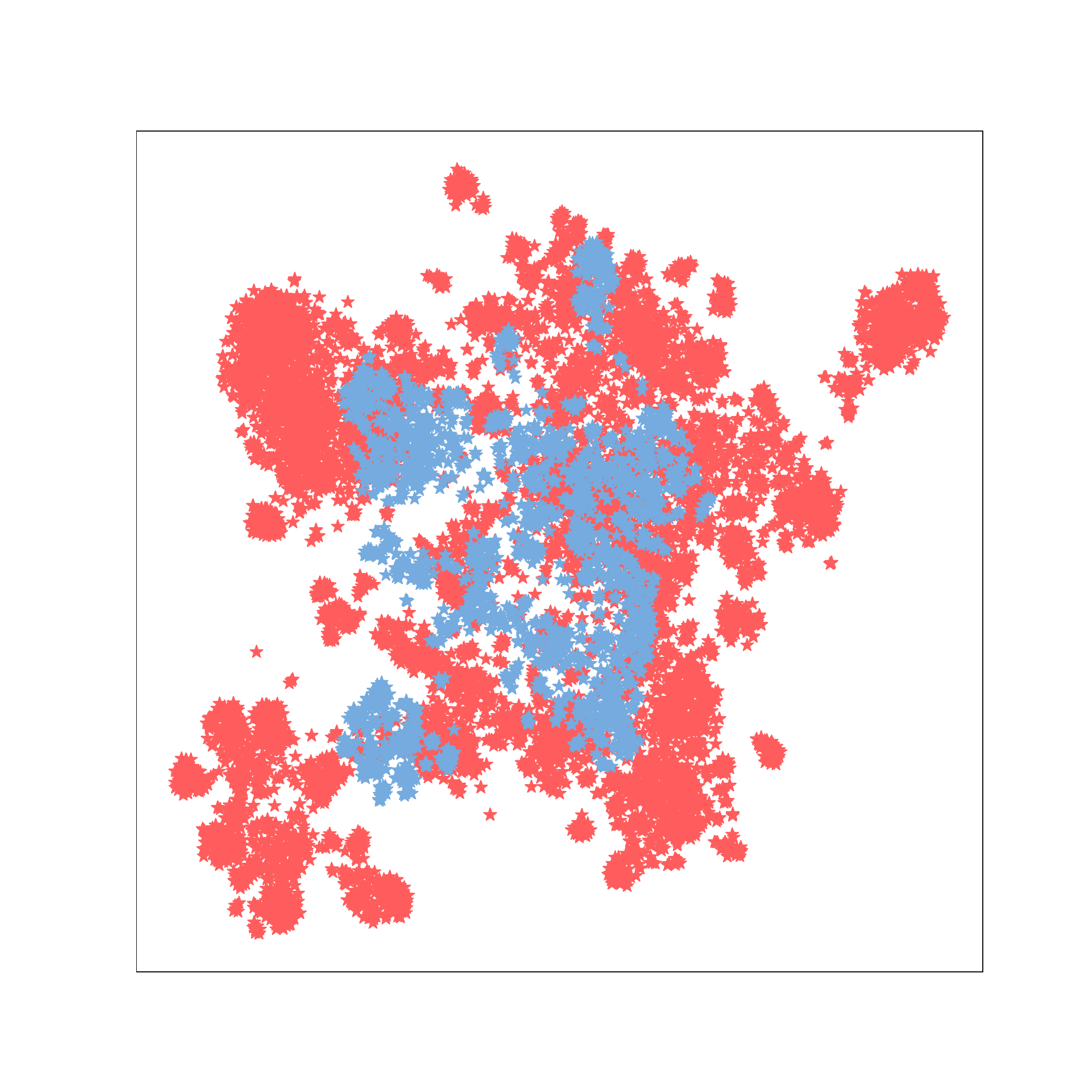}
    }
    \subfigure[MMSSL]{
   \centering
    \includegraphics[width=0.25\linewidth]{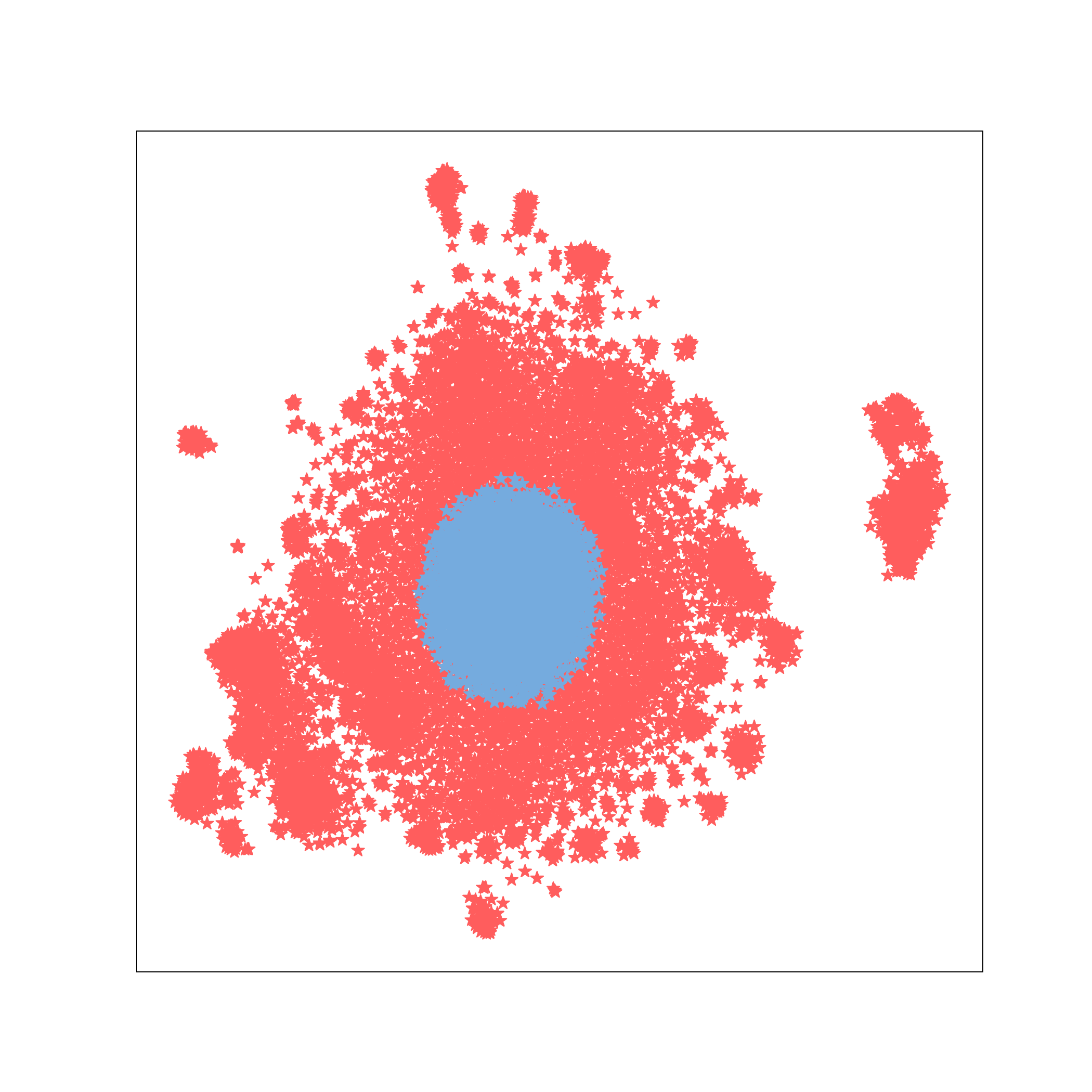}
    }
    \\
    \vspace{-0.35cm}
    \centering
    \subfigure[MKGAT]{
       \centering
        \includegraphics[width=0.25\linewidth]{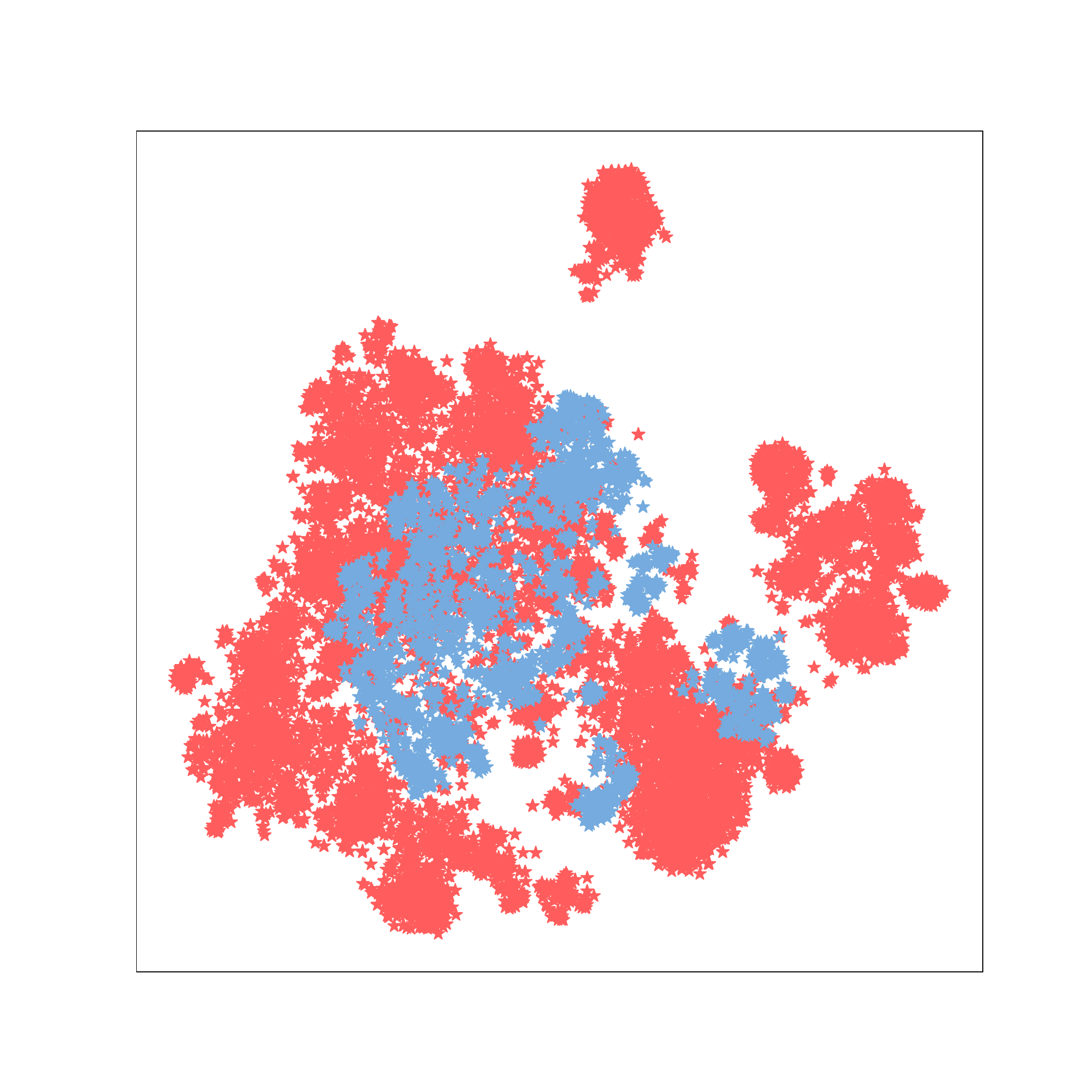}
    }
    \subfigure[DropoutNet]{
   \centering
    \includegraphics[width=0.25\linewidth]{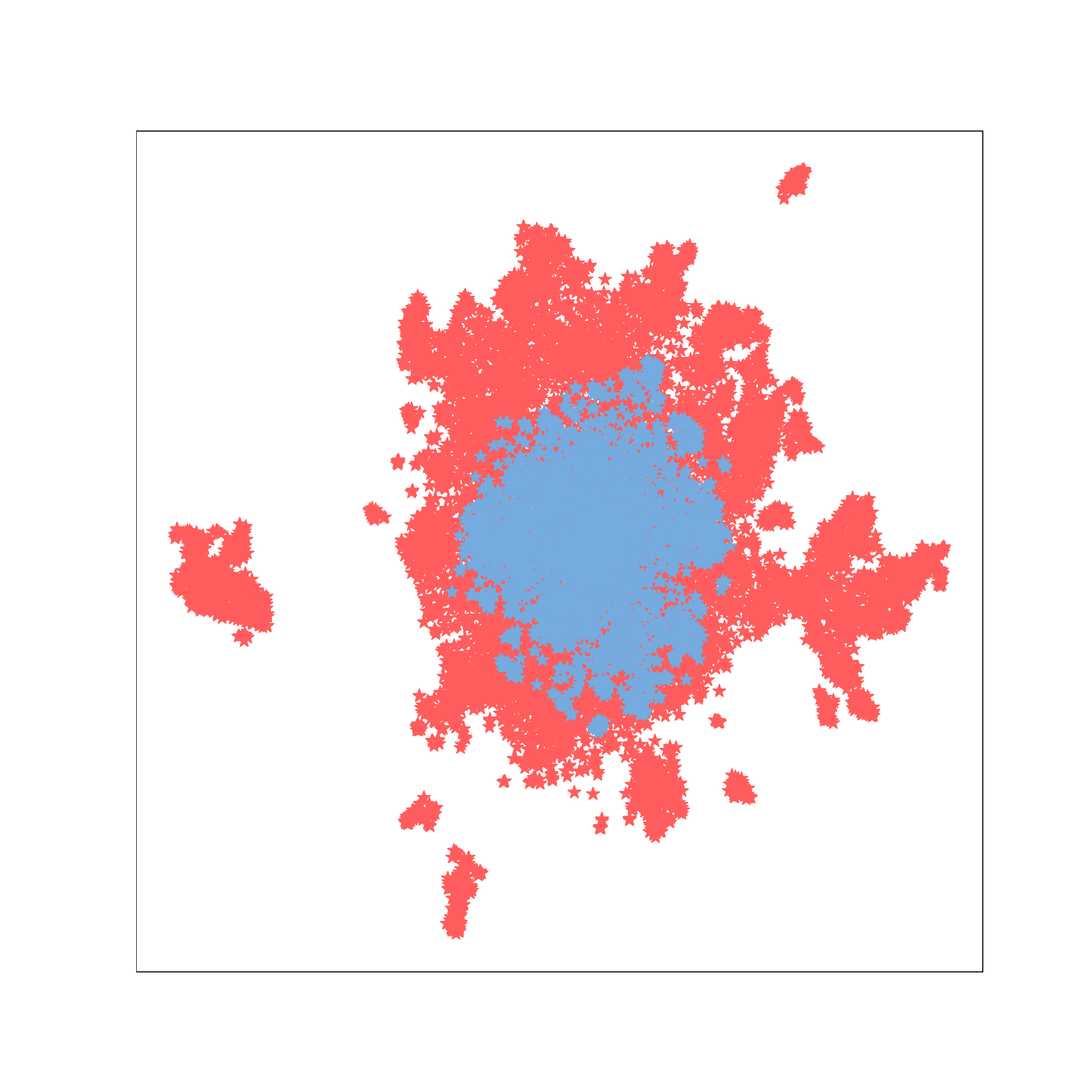}
    }
    \subfigure[\textbf{Firzen (Ours)}]{
   \centering
    \includegraphics[width=0.25\linewidth]{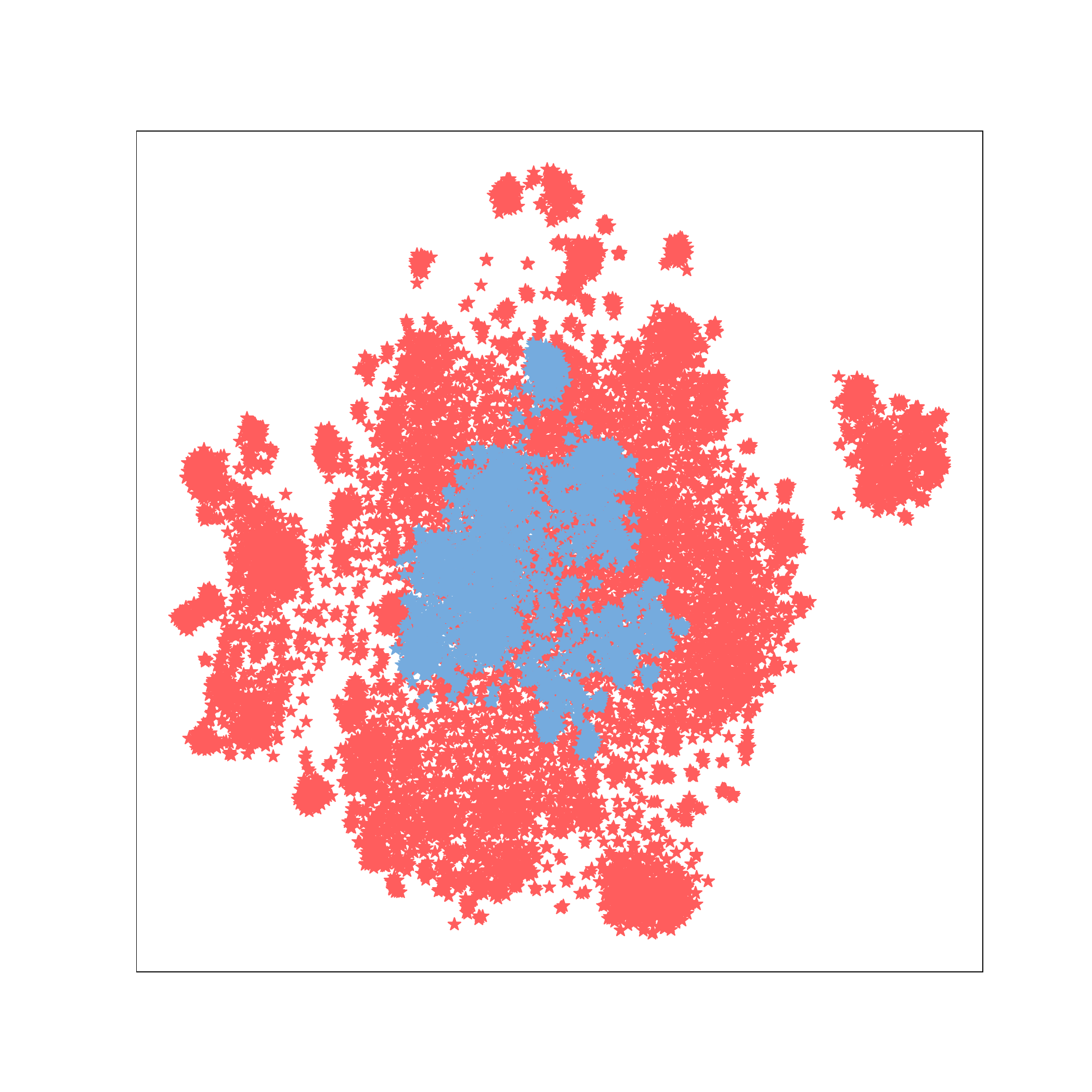}
    }
}\caption{t-SNE visualization of strict cold-start (blue) and warm-start (red) item embeddings' distribution.\label{fig:t-sne}}
\end{figure}

Finally, we visualize the embeddings of strict cold-start and warm-start items to illustrate the effectiveness of Firzen on both recommendation scenarios. Specifically, we visualize the distribution of the cold and warm item embeddings by reducing their dimension to two with t-SNE\cite{van2008visualizing}. Then we compare the distribution of the typical baselines LightGCN, KGAT, MMSSL, MKGAT, DropoutNet and Firzen. Observing from Fig. \ref{fig:t-sne}, the strict cold-start embeddings produced by LightGCN and MMSSL are compactly distributed and significantly different from the warm embeddings. Thus, as reported in Table \ref{tab:main_results} and \ref{tab:Weixin_results}, the strict cold-start recommendation performance is limited. For KGAT, MKGAT and DropoutNet, the distribution of cold embeddings is relatively improved while that of warm embeddings is affected, resulting in degraded performance in warm-start scenario. Benefiting from propagating BA, KA and MA features from warm-start to strict cold-start items, the cold embeddings of Firzen have more similar distribution as the warm embeddings, while preserving the raw distribution of warm embeddings like LightGCN and MMSSL. 

\section{Related Work}
\subsection{Cold-start Item Recommendation}

The cold-start item recommendation problem is a long-standing challenge in recommender systems. The main line of research lies in the content-based approaches, which utilizes side information of items to overcome the problem of sparse data. Some works model item content directly \cite{ebesu2017neural}. Some works learn the correlation of id embeddings and side information, and learn a generator to project the item content into the warm item embedding space\cite{gantner2010learning,mo2015image,barkan2019cb2cf,pan2019warm,sun2020lara,chen2022generative}. Another group of works tackle the problem from the perspective of robust learning. It treats cold-start items as warm-start items missing interactions, and attempt to infer the warm embeddings for them \cite{volkovs2017dropoutnet, du2020learn}. For instance, DropoutNet \cite{volkovs2017dropoutnet} randomly samples a subset of users/items and remove their behavior-based features to simulate the cold-start scenario during the training phase. Besides, some efforts use meta-learning to address the cold-start recommendation \cite{lee2019melu, zhu2021learning}. For instance, MWUF\cite{zhu2021learning} generates warm embeddings for cold-start items based on their features and ID embeddings using meta shifting network and meta scaling network.

\subsection{Multi-modal Recommendation}

Multi-modal recommendation aims to utilize the rich multi-modal content information (e.g., texts, images) of items in the recommendation. Most early multi-modal recommendation models utilize deep learning techniques to explore users' preferences of different modalities on top of collaborative filtering paradigm \cite{he2016vbpr, chen2019personalized, liu2019user}. Recently, another group of research introduces GNNs into the multi-modal recommendation systems to boost the performance \cite{wei2019mmgcn,wang2021dualgnn,zhang2021mining,zhou2022tale}.
For example, LATTICE\cite{zhang2021mining} builds and dynamically updates item-item relation graphs, while FREEDOM\cite{zhou2022tale} further finds that the learning of item-item graphs are negligible and freezes the item-item graph to improve the efficiency. 

\subsection{Knowledge-aware Recommendation}

Knowledge-aware recommendation aims at enhance the representations of users and items by introducing external KGs, which can be roughly grouped into two categories: path-based methods and embedding-based methods.  Path-based methods\cite{hu2018leveraging, wang2018ripplenet, wang2019knowledge, wang2019kgat, xia2021knowledge},  aim to explore the potential connect between items in KG by constructing meta-path for information propagation. For example, MCRec\cite{hu2018leveraging} proposes meta-path-based mutual attention mechanism, which produces user, item and meta-path-based context representations. Embedding-based methods\cite{zhang2016collaborative, wang2018dkn,xin2019relational, tian2021joint, wang2022tower, chen2022attentive} integrate the representation learning of KGs to enhance the user and item embeddings. For instance, the embedding of items' structural knowledge is encoded with TransR\cite{lin2015learning} for CKE\cite{zhang2016collaborative}. 

\section{Conclusion}
In this work, we focus on the problem of building competitive models for both strict cold-start and warm-start item recommendation by proposing a model collaborating multi-modal content and KGs named Firzen. In Firzen, Side information-Aware Heterogeneous Graph Learning augments the representation by behavior-aware, knowledge-aware and modality-aware user-item information propagation. Moreover, Modality-Specific Homogeneous Graph Learning passes the message based on potential connection among strict cold-start and warm-start items, alleviating the strict cold-start problem while preserving the performance of warm-start recommendation. Extensive experiments on four benchmarks built upon Amazon datasets and a real-world industrial dataset Weixin-Sports verify the effectiveness of the proposed Firzen on achieving competitive performance in both strict cold-start and warm-start scenarios compared with existing methods. 

\section*{Acknowledgement}
This work was supported by the grants from the National Natural Science Foundation of China (61925201, 62132001, 62272013, U22B2048) and by 2022 Tencent Wechat Rhino-Bird Focused Research Program.
\bibliographystyle{IEEEtran}
\bibliography{references}

\begin{thebibliography}{10}
\providecommand{\url}[1]{#1}
\csname url@samestyle\endcsname
\providecommand{\newblock}{\relax}
\providecommand{\bibinfo}[2]{#2}
\providecommand{\BIBentrySTDinterwordspacing}{\spaceskip=0pt\relax}
\providecommand{\BIBentryALTinterwordstretchfactor}{4}
\providecommand{\BIBentryALTinterwordspacing}{\spaceskip=\fontdimen2\font plus
\BIBentryALTinterwordstretchfactor\fontdimen3\font minus \fontdimen4\font\relax}
\providecommand{\BIBforeignlanguage}[2]{{%
\expandafter\ifx\csname l@#1\endcsname\relax
\typeout{** WARNING: IEEEtran.bst: No hyphenation pattern has been}%
\typeout{** loaded for the language `#1'. Using the pattern for}%
\typeout{** the default language instead.}%
\else
\language=\csname l@#1\endcsname
\fi
#2}}
\providecommand{\BIBdecl}{\relax}
\BIBdecl

\bibitem{liu2019uservideo}
S.~Liu, Z.~Chen, H.~Liu, and X.~Hu, ``User-video co-attention network for personalized micro-video recommendation,'' in \emph{The world wide web conference}, 2019, pp. 3020--3026.

\bibitem{gharibshah2021user}
Z.~Gharibshah and X.~Zhu, ``User response prediction in online advertising,'' \emph{aCM Computing Surveys (CSUR)}, vol.~54, no.~3, pp. 1--43, 2021.

\bibitem{wang2020time}
J.~Wang, R.~Louca, D.~Hu, C.~Cellier, J.~Caverlee, and L.~Hong, ``Time to shop for valentine's day: Shopping occasions and sequential recommendation in e-commerce,'' in \emph{Proceedings of the 13th International Conference on Web Search and Data Mining}, 2020, pp. 645--653.

\bibitem{koren2009matrix}
Y.~Koren, R.~Bell, and C.~Volinsky, ``Matrix factorization techniques for recommender systems,'' \emph{Computer}, vol.~42, no.~8, pp. 30--37, 2009.

\bibitem{rendle2009bpr}
S.~Rendle, C.~Freudenthaler, Z.~Gantner, and L.~Schmidt-Thieme, ``Bpr: Bayesian personalized ranking from implicit feedback,'' in \emph{Proceedings of the Twenty-Fifth Conference on Uncertainty in Artificial Intelligence}, 2009, pp. 452--461.

\bibitem{he2020lightgcn}
X.~He, K.~Deng, X.~Wang, Y.~Li, Y.~Zhang, and M.~Wang, ``Lightgcn: Simplifying and powering graph convolution network for recommendation,'' in \emph{Proceedings of the 43rd International ACM SIGIR conference on research and development in Information Retrieval}, 2020, pp. 639--648.

\bibitem{yuan2022tenrec}
G.~Yuan, F.~Yuan, Y.~Li, B.~Kong, S.~Li, L.~Chen, M.~Yang, C.~Yu, B.~Hu, Z.~Li \emph{et~al.}, ``Tenrec: A large-scale multipurpose benchmark dataset for recommender systems,'' \emph{Advances in Neural Information Processing Systems}, vol.~35, pp. 11\,480--11\,493, 2022.

\bibitem{wei2019mmgcn}
Y.~Wei, X.~Wang, L.~Nie, X.~He, R.~Hong, and T.-S. Chua, ``Mmgcn: Multi-modal graph convolution network for personalized recommendation of micro-video,'' in \emph{Proceedings of the 27th ACM international conference on multimedia}, 2019, pp. 1437--1445.

\bibitem{zhou2023bootstrap}
X.~Zhou, H.~Zhou, Y.~Liu, Z.~Zeng, C.~Miao, P.~Wang, Y.~You, and F.~Jiang, ``Bootstrap latent representations for multi-modal recommendation,'' in \emph{Proceedings of the ACM Web Conference 2023}, 2023, pp. 845--854.

\bibitem{wei2023multi}
W.~Wei, C.~Huang, L.~Xia, and C.~Zhang, ``Multi-modal self-supervised learning for recommendation,'' in \emph{Proceedings of the ACM Web Conference 2023}, 2023, pp. 790--800.

\bibitem{wang2019kgat}
X.~Wang, X.~He, Y.~Cao, M.~Liu, and T.-S. Chua, ``Kgat: Knowledge graph attention network for recommendation,'' in \emph{Proceedings of the 25th ACM SIGKDD international conference on knowledge discovery \& data mining}, 2019, pp. 950--958.

\bibitem{xian2019reinforcement}
Y.~Xian, Z.~Fu, S.~Muthukrishnan, G.~De~Melo, and Y.~Zhang, ``Reinforcement knowledge graph reasoning for explainable recommendation,'' in \emph{Proceedings of the 42nd international ACM SIGIR conference on research and development in information retrieval}, 2019, pp. 285--294.

\bibitem{huang2021knowledge}
C.~Huang, H.~Xu, Y.~Xu, P.~Dai, L.~Xia, M.~Lu, L.~Bo, H.~Xing, X.~Lai, and Y.~Ye, ``Knowledge-aware coupled graph neural network for social recommendation,'' in \emph{Proceedings of the AAAI conference on artificial intelligence}, vol.~35, no.~5, 2021, pp. 4115--4122.

\bibitem{zhao2017meta}
H.~Zhao, Q.~Yao, J.~Li, Y.~Song, and D.~L. Lee, ``Meta-graph based recommendation fusion over heterogeneous information networks,'' in \emph{Proceedings of the 23rd ACM SIGKDD international conference on knowledge discovery and data mining}, 2017, pp. 635--644.

\bibitem{wang2018ripplenet}
H.~Wang, F.~Zhang, J.~Wang, M.~Zhao, W.~Li, X.~Xie, and M.~Guo, ``Ripplenet: Propagating user preferences on the knowledge graph for recommender systems,'' in \emph{Proceedings of the 27th ACM international conference on information and knowledge management}, 2018, pp. 417--426.

\bibitem{wang2019explainable}
X.~Wang, D.~Wang, C.~Xu, X.~He, Y.~Cao, and T.-S. Chua, ``Explainable reasoning over knowledge graphs for recommendation,'' in \emph{Proceedings of the AAAI conference on artificial intelligence}, vol.~33, no.~01, 2019, pp. 5329--5336.

\bibitem{tai2020mvin}
C.-Y. Tai, M.-R. Wu, Y.-W. Chu, S.-Y. Chu, and L.-W. Ku, ``Mvin: Learning multiview items for recommendation,'' in \emph{Proceedings of the 43rd international ACM SIGIR conference on research and development in information retrieval}, 2020, pp. 99--108.

\bibitem{wang2021learning}
X.~Wang, T.~Huang, D.~Wang, Y.~Yuan, Z.~Liu, X.~He, and T.-S. Chua, ``Learning intents behind interactions with knowledge graph for recommendation,'' in \emph{Proceedings of the web conference 2021}, 2021, pp. 878--887.

\bibitem{xia2021knowledge}
L.~Xia, C.~Huang, Y.~Xu, P.~Dai, X.~Zhang, H.~Yang, J.~Pei, and L.~Bo, ``Knowledge-enhanced hierarchical graph transformer network for multi-behavior recommendation,'' in \emph{Proceedings of the AAAI Conference on Artificial Intelligence}, vol.~35, no.~5, 2021, pp. 4486--4493.

\bibitem{sun2020multi}
R.~Sun, X.~Cao, Y.~Zhao, J.~Wan, K.~Zhou, F.~Zhang, Z.~Wang, and K.~Zheng, ``Multi-modal knowledge graphs for recommender systems,'' in \emph{Proceedings of the 29th ACM international conference on information \& knowledge management}, 2020, pp. 1405--1414.

\bibitem{volkovs2017dropoutnet}
M.~Volkovs, G.~Yu, and T.~Poutanen, ``Dropoutnet: Addressing cold start in recommender systems,'' \emph{Advances in neural information processing systems}, vol.~30, 2017.

\bibitem{zhang2021mining}
J.~Zhang, Y.~Zhu, Q.~Liu, S.~Wu, S.~Wang, and L.~Wang, ``Mining latent structures for multimedia recommendation,'' in \emph{Proceedings of the 29th ACM International Conference on Multimedia}, 2021, pp. 3872--3880.

\bibitem{qian2020attribute}
T.~Qian, Y.~Liang, Q.~Li, and H.~Xiong, ``Attribute graph neural networks for strict cold start recommendation,'' \emph{IEEE Transactions on Knowledge and Data Engineering}, vol.~34, no.~8, pp. 3597--3610, 2020.

\bibitem{zhou2022tale}
X.~Zhou, ``A tale of two graphs: Freezing and denoising graph structures for multimodal recommendation,'' \emph{arXiv preprint arXiv:2211.06924}, 2022.

\bibitem{zhou2023enhancing}
H.~Zhou, X.~Zhou, and Z.~Shen, ``Enhancing dyadic relations with homogeneous graphs for multimodal recommendation,'' \emph{arXiv preprint arXiv:2301.12097}, 2023.

\bibitem{chen2009fast}
J.~Chen, H.-r. Fang, and Y.~Saad, ``Fast approximate knn graph construction for high dimensional data via recursive lanczos bisection.'' \emph{Journal of Machine Learning Research}, vol.~10, no.~9, 2009.

\bibitem{hinton2012improving}
G.~E. Hinton, N.~Srivastava, A.~Krizhevsky, I.~Sutskever, and R.~R. Salakhutdinov, ``Improving neural networks by preventing co-adaptation of feature detectors,'' \emph{arXiv preprint arXiv:1207.0580}, 2012.

\bibitem{qiu2018deepinf}
J.~Qiu, J.~Tang, H.~Ma, Y.~Dong, K.~Wang, and J.~Tang, ``Deepinf: Social influence prediction with deep learning,'' in \emph{Proceedings of the 24th ACM SIGKDD international conference on knowledge discovery \& data mining}, 2018, pp. 2110--2119.

\bibitem{vaswani2017attention}
A.~Vaswani, N.~Shazeer, N.~Parmar, J.~Uszkoreit, L.~Jones, A.~N. Gomez, {\L}.~Kaiser, and I.~Polosukhin, ``Attention is all you need,'' \emph{Advances in neural information processing systems}, vol.~30, 2017.

\bibitem{jang2017categorical}
E.~Jang, S.~Gu, and B.~Poole, ``Categorical reparametrization with gumble-softmax,'' in \emph{International Conference on Learning Representations (ICLR 2017)}.\hskip 1em plus 0.5em minus 0.4em\relax OpenReview. net, 2017.

\bibitem{gulrajani2017improved}
I.~Gulrajani, F.~Ahmed, M.~Arjovsky, V.~Dumoulin, and A.~C. Courville, ``Improved training of wasserstein gans,'' \emph{Advances in neural information processing systems}, vol.~30, 2017.

\bibitem{he2020momentum}
K.~He, H.~Fan, Y.~Wu, S.~Xie, and R.~Girshick, ``Momentum contrast for unsupervised visual representation learning,'' in \emph{Proceedings of the IEEE/CVF conference on computer vision and pattern recognition}, 2020, pp. 9729--9738.

\bibitem{lin2015learning}
Y.~Lin, Z.~Liu, M.~Sun, Y.~Liu, and X.~Zhu, ``Learning entity and relation embeddings for knowledge graph completion,'' in \emph{Proceedings of the AAAI conference on artificial intelligence}, vol.~29, no.~1, 2015.

\bibitem{kingma2015adam}
D.~P. Kingma and J.~L. Ba, ``Adam: A method for stochastic gradient descent,'' in \emph{ICLR: international conference on learning representations}.\hskip 1em plus 0.5em minus 0.4em\relax ICLR US., 2015, pp. 1--15.

\bibitem{mcauley2015image}
J.~McAuley, C.~Targett, Q.~Shi, and A.~Van Den~Hengel, ``Image-based recommendations on styles and substitutes,'' in \emph{Proceedings of the 38th international ACM SIGIR conference on research and development in information retrieval}, 2015, pp. 43--52.

\bibitem{he2016ups}
R.~He and J.~McAuley, ``Ups and downs: Modeling the visual evolution of fashion trends with one-class collaborative filtering,'' in \emph{proceedings of the 25th international conference on world wide web}, 2016, pp. 507--517.

\bibitem{ni2019justifying}
J.~Ni, J.~Li, and J.~McAuley, ``Justifying recommendations using distantly-labeled reviews and fine-grained aspects,'' in \emph{Proceedings of the 2019 conference on empirical methods in natural language processing and the 9th international joint conference on natural language processing (EMNLP-IJCNLP)}, 2019, pp. 188--197.

\bibitem{reimers2019sentence}
N.~Reimers and I.~Gurevych, ``Sentence-bert: Sentence embeddings using siamese bert-networks,'' \emph{arXiv preprint arXiv:1908.10084}, 2019.

\bibitem{wei2021contrastive}
Y.~Wei, X.~Wang, Q.~Li, L.~Nie, Y.~Li, X.~Li, and T.-S. Chua, ``Contrastive learning for cold-start recommendation,'' in \emph{Proceedings of the 29th ACM International Conference on Multimedia}, 2021, pp. 5382--5390.

\bibitem{sparck1972statistical}
K.~Sparck~Jones, ``A statistical interpretation of term specificity and its application in retrieval,'' \emph{Journal of documentation}, vol.~28, no.~1, pp. 11--21, 1972.

\bibitem{wu2021self}
J.~Wu, X.~Wang, F.~Feng, X.~He, L.~Chen, J.~Lian, and X.~Xie, ``Self-supervised graph learning for recommendation,'' in \emph{Proceedings of the 44th international ACM SIGIR conference on research and development in information retrieval}, 2021, pp. 726--735.

\bibitem{mao2021simplex}
K.~Mao, J.~Zhu, J.~Wang, Q.~Dai, Z.~Dong, X.~Xiao, and X.~He, ``Simplex: A simple and strong baseline for collaborative filtering,'' in \emph{Proceedings of the 30th ACM International Conference on Information \& Knowledge Management}, 2021, pp. 1243--1252.

\bibitem{zhang2016collaborative}
F.~Zhang, N.~J. Yuan, D.~Lian, X.~Xie, and W.-Y. Ma, ``Collaborative knowledge base embedding for recommender systems,'' in \emph{Proceedings of the 22nd ACM SIGKDD international conference on knowledge discovery and data mining}, 2016, pp. 353--362.

\bibitem{wang2019knowledge}
H.~Wang, M.~Zhao, X.~Xie, W.~Li, and M.~Guo, ``Knowledge graph convolutional networks for recommender systems,'' in \emph{The world wide web conference}, 2019, pp. 3307--3313.

\bibitem{kgnnls}
H.~Wang, F.~Zhang, M.~Zhang, J.~Leskovec, M.~Zhao, W.~Li, and Z.~Wang, ``Knowledge-aware graph neural networks with label smoothness regularization for recommender systems,'' in \emph{Proceedings of the 25th ACM SIGKDD international conference on knowledge discovery \& data mining}, 2019, pp. 968--977.

\bibitem{he2016vbpr}
R.~He and J.~McAuley, ``Vbpr: visual bayesian personalized ranking from implicit feedback,'' in \emph{Proceedings of the AAAI conference on artificial intelligence}, vol.~30, no.~1, 2016.

\bibitem{zhou2023contrastive}
Z.~Zhou, L.~Zhang, and N.~Yang, ``Contrastive collaborative filtering for cold-start item recommendation,'' in \emph{Proceedings of the ACM Web Conference 2023}, 2023, pp. 928--937.

\bibitem{van2008visualizing}
L.~Van~der Maaten and G.~Hinton, ``Visualizing data using t-sne.'' \emph{Journal of machine learning research}, vol.~9, no.~11, 2008.

\bibitem{ebesu2017neural}
T.~Ebesu and Y.~Fang, ``Neural semantic personalized ranking for item cold-start recommendation,'' \emph{Information Retrieval Journal}, vol.~20, pp. 109--131, 2017.

\bibitem{gantner2010learning}
Z.~Gantner, L.~Drumond, C.~Freudenthaler, S.~Rendle, and L.~Schmidt-Thieme, ``Learning attribute-to-feature mappings for cold-start recommendations,'' in \emph{2010 IEEE International Conference on Data Mining}.\hskip 1em plus 0.5em minus 0.4em\relax IEEE, 2010, pp. 176--185.

\bibitem{mo2015image}
K.~Mo, B.~Liu, L.~Xiao, Y.~Li, and J.~Jiang, ``Image feature learning for cold start problem in display advertising,'' in \emph{Twenty-Fourth International Joint Conference on Artificial Intelligence}, 2015.

\bibitem{barkan2019cb2cf}
O.~Barkan, N.~Koenigstein, E.~Yogev, and O.~Katz, ``Cb2cf: a neural multiview content-to-collaborative filtering model for completely cold item recommendations,'' in \emph{Proceedings of the 13th ACM Conference on Recommender Systems}, 2019, pp. 228--236.

\bibitem{pan2019warm}
F.~Pan, S.~Li, X.~Ao, P.~Tang, and Q.~He, ``Warm up cold-start advertisements: Improving ctr predictions via learning to learn id embeddings,'' in \emph{Proceedings of the 42nd International ACM SIGIR Conference on Research and Development in Information Retrieval}, 2019, pp. 695--704.

\bibitem{sun2020lara}
C.~Sun, H.~Liu, M.~Liu, Z.~Ren, T.~Gan, and L.~Nie, ``Lara: Attribute-to-feature adversarial learning for new-item recommendation,'' in \emph{Proceedings of the 13th international conference on web search and data mining}, 2020, pp. 582--590.

\bibitem{chen2022generative}
H.~Chen, Z.~Wang, F.~Huang, X.~Huang, Y.~Xu, Y.~Lin, P.~He, and Z.~Li, ``Generative adversarial framework for cold-start item recommendation,'' in \emph{Proceedings of the 45th International ACM SIGIR Conference on Research and Development in Information Retrieval}, 2022, pp. 2565--2571.

\bibitem{du2020learn}
X.~Du, X.~Wang, X.~He, Z.~Li, J.~Tang, and T.-S. Chua, ``How to learn item representation for cold-start multimedia recommendation?'' in \emph{Proceedings of the 28th ACM International Conference on Multimedia}, 2020, pp. 3469--3477.

\bibitem{lee2019melu}
H.~Lee, J.~Im, S.~Jang, H.~Cho, and S.~Chung, ``Melu: Meta-learned user preference estimator for cold-start recommendation,'' in \emph{Proceedings of the 25th ACM SIGKDD International Conference on Knowledge Discovery \& Data Mining}, 2019, pp. 1073--1082.

\bibitem{zhu2021learning}
Y.~Zhu, R.~Xie, F.~Zhuang, K.~Ge, Y.~Sun, X.~Zhang, L.~Lin, and J.~Cao, ``Learning to warm up cold item embeddings for cold-start recommendation with meta scaling and shifting networks,'' in \emph{Proceedings of the 44th International ACM SIGIR Conference on Research and Development in Information Retrieval}, 2021, pp. 1167--1176.

\bibitem{chen2019personalized}
X.~Chen, H.~Chen, H.~Xu, Y.~Zhang, Y.~Cao, Z.~Qin, and H.~Zha, ``Personalized fashion recommendation with visual explanations based on multimodal attention network: Towards visually explainable recommendation,'' in \emph{Proceedings of the 42nd International ACM SIGIR Conference on Research and Development in Information Retrieval}, 2019, pp. 765--774.

\bibitem{liu2019user}
F.~Liu, Z.~Cheng, C.~Sun, Y.~Wang, L.~Nie, and M.~Kankanhalli, ``User diverse preference modeling by multimodal attentive metric learning,'' in \emph{Proceedings of the 27th ACM international conference on multimedia}, 2019, pp. 1526--1534.

\bibitem{wang2021dualgnn}
Q.~Wang, Y.~Wei, J.~Yin, J.~Wu, X.~Song, and L.~Nie, ``Dualgnn: Dual graph neural network for multimedia recommendation,'' \emph{IEEE Transactions on Multimedia}, 2021.

\bibitem{hu2018leveraging}
B.~Hu, C.~Shi, W.~X. Zhao, and P.~S. Yu, ``Leveraging meta-path based context for top-n recommendation with a neural co-attention model,'' in \emph{Proceedings of the 24th ACM SIGKDD international conference on knowledge discovery \& data mining}, 2018, pp. 1531--1540.

\bibitem{wang2018dkn}
H.~Wang, F.~Zhang, X.~Xie, and M.~Guo, ``Dkn: Deep knowledge-aware network for news recommendation,'' in \emph{Proceedings of the 2018 world wide web conference}, 2018, pp. 1835--1844.

\bibitem{xin2019relational}
X.~Xin, X.~He, Y.~Zhang, Y.~Zhang, and J.~Jose, ``Relational collaborative filtering: Modeling multiple item relations for recommendation,'' in \emph{Proceedings of the 42nd international ACM SIGIR conference on research and development in information retrieval}, 2019, pp. 125--134.

\bibitem{tian2021joint}
Y.~Tian, Y.~Yang, X.~Ren, P.~Wang, F.~Wu, Q.~Wang, and C.~Li, ``Joint knowledge pruning and recurrent graph convolution for news recommendation,'' in \emph{Proceedings of the 44th International ACM SIGIR Conference on Research and Development in Information Retrieval}, 2021, pp. 51--60.

\bibitem{wang2022tower}
S.~Wang, H.~Li, C.~C. Cao, X.-H. Li, N.~N. Fai, J.~Liu, X.~Xue, H.~Song, J.~Li, G.~Gu \emph{et~al.}, ``Tower bridge net (tb-net): Bidirectional knowledge graph aware embedding propagation for explainable recommender systems,'' in \emph{2022 IEEE 38th International Conference on Data Engineering (ICDE)}.\hskip 1em plus 0.5em minus 0.4em\relax IEEE, 2022, pp. 3268--3279.

\bibitem{chen2022attentive}
Y.~Chen, Y.~Yang, Y.~Wang, J.~Bai, X.~Song, and I.~King, ``Attentive knowledge-aware graph convolutional networks with collaborative guidance for personalized recommendation,'' in \emph{2022 IEEE 38th International Conference on Data Engineering (ICDE)}.\hskip 1em plus 0.5em minus 0.4em\relax IEEE, 2022, pp. 299--311.

\end{thebibliography}

\end{document}